\documentclass[%
 reprint,
 amsmath,amssymb,
 aps,
pra,
floatfix,
]{revtex4-2}

\usepackage{graphicx}% Include figure files
\usepackage{dcolumn}% Align table columns on decimal point
\usepackage{bm}% bold math
\usepackage{xcolor}
\usepackage{hyperref}% add hypertext capabilities
\usepackage{multirow}

\def\expect#1{{\langle #1\rangle}}
\def\be{\begin{eqnarray}}
\def\ee{\end{eqnarray}}
\def\ket#1{\left|#1\right\rangle}
\def\bra#1{\left\langle#1\right|}

\def\Edit#1{\color{black} #1}

\usepackage[normalem]{ulem}   % provides \sout (strike-out); normalem keeps \emph behaving normally
\newif\ifshowstrikes
\showstrikesfalse  % <<< change this to \showstrikesfalse for the final version

\makeatletter
\newcommand{\strike}[1]{%
  \ifshowstrikes
    {\color{red}\sout{#1}}% %show deletions in red
  \else
  \fi}
\makeatother

\begin{document}

\title{Perspective: Practical Atom-Based Quantum Sensors}% Force line breaks with \\

\author{Justin M. Brown}
%\email{brown@isi.edu}

\altaffiliation[Present Address: ]{The Charles Stark Draper Laboratory, Inc., Cambridge, MA 02139, USA}
\email{jbrown@draper.com}
\affiliation{Information Sciences Institute, University of Southern California, Waltham, MA 02451, USA
}
\author{Thad G. Walker}
\email{tgwalker@wisc.edu}
\affiliation{Department of Physics, University of Wisconsin-Madison
}%

\date{\today}% It is always \today, today,

\begin{abstract}
%Quantum sensors based on atoms offer the potential to perform measurements with the highest precision as well as the best long-term stability.  This unique advantage of the atomic measurement platform originates from the property that all atoms of a given species are indistinguishable and the relative simplicity with which quantum properties can be fully exploited in state preparation and detection using modern laser and electo-optical tools.  As such, atomic clocks have long been the established frequency standard and best reference to timekeeping. More recently, electromagnetic and inertial sensors are maturing towards one day systematically out-performing classical systems. 
%Atomic vapors, manipulated and probed by light and other electromagnetic fields, constitute  versatile and powerful quantum systems for sensing applications.  In particular, quantum sensors based on atoms offer the potential to perform measurements with the highest precision or long-term stability.  This unique advantage originates from the property that all atoms of a given species are indistinguishable and the relative simplicity with which quantum properties can be exploited in state preparation and detection using modern laser and electro-optic tools.  

Atomic vapors, manipulated and probed by light and other electromagnetic fields, constitute  versatile and powerful quantum systems for sensing applications. Atoms are \strike{not only \strike{indistiguishable} {\Edit identical}, they are} {\Edit identical, }isolatable, interfaceable, and intelligible.
%, and thus inherently "Sens-ible" HAHA. 
These features, coupled with the relative simplicity with which quantum properties can be exploited in state preparation and detection using modern laser and electro-optic tools, make atoms very attractive for sensing applications.  This Perspective discusses the potential and process for realizing practical quantum sensors using atoms.

%{\Edit In this perspective, we describe a generalized quantum sensor, highlighting atomic properties and key strategies required to extract the full measurement performance capability in the quantum system.  A recurring theme is the importance of deep understanding of basic atomic physics processes in order to fully exploit the potential for quantum measurement.  Such approaches offer the potential for rapid realization of mature sensors that positively impact numerous applications in position, navigation, and timing (PNT), biomedical diagnostics, and communications.}
\end{abstract}

\maketitle

\tableofcontents

\section{Atom-Based Quantum Sensors}

Quantum sensors based on atoms take advantage of the inherent stability, reproducibility, and sensitivity of atomic spectroscopy to make measurements of physical quantities of interest.  As ``interests" may vary, there is a measure of subjectivity involved ranging from the hypothetical ({\Edit i.e.,} the existence of axion-like quantum fields for fundamental science \cite{Safronova2018}) to the practical ({\Edit i.e.,} precise timing signals on satellites using atomic clocks to enable GPS navigation \cite{Camparo2023AppliedSpace,DupuisRyan20082008Symposium}). Atom-based quantum sensors typically use photon-mediated interactions to probe quantum states of vapor phase atoms whose energies are sensitive to the property of interest.  The sensor may seek to detect the oscillation frequency between isolated energy levels to serve as a frequency reference for a clock, or track the frequency shift in those levels in response to an environmental magnetic field as a magnetometer.  The sensor may also seek to detect classical information (encoded in modulations of an electromagnetic  field), or read out the quantum state of another atom (inside a quantum information processor or clock).

For applications beyond those of tests of fundamental physics, the practical sensor user is not necessarily interested in the ultimate sensitivity that can be achieved in a research laboratory, but in reliably using the sensor as a tool in an uncontrolled environment subject to perturbations, where ease of use benefits from a manageable size, weight, and power.  The operator may even be a non-specialist in the inner workings of the sensor, necessitating turn-key operation.  To manage these constraints, the sensor may be willfully designed to operate away from the achievable quantum measurement limits while still taking advantage of the high sensitivity, stability, and reproducibility inherent to atomic quantum systems.  

The development of such ``practical" sensors, the focus of this Perspective, is enabled by decades of science experiments devoted to achieving maximum sensitivity in the pursuit of studies of fundamental physics.  Indeed, most atomic sensor technologies should be properly viewed as descendants or offshoots of laboratory precision measurements.  Precision measurements using atomic, molecular, and optical (AMO) techniques have been thoroughly reviewed recently \cite{Safronova2018} and anyone interested in ``practical" atomic sensors would be well-advised to become familiar with the methods being pursued in forefront precision measurement experiments \cite{Safronova2021QuantumSensors,DeMille2017,Indelicato2019QEDIons}. % {\Edit One should not confuse such fundamental experiments as ``impractical"; use here is to explicitly distinguish a specific set of use cases.}  
Similarly, the rapidly developing field of quantum information science (QIS) can be considered as sensing of quantum information \cite{Kaufman2021QuantumMolecules}. 
 Unquestionably, the systems developed for these applications are high-quality atomic quantum sensors in their own right, but their complexity puts them outside the scope of our interests here.
% we have chosen to omit this specialized topic from the considerations of this paper .

Early atomic quantum sensors, atomic clocks in particular, grew out of research on atomic beam magnetic resonance and often utilized Stern-Gerlach magnets, microwave cavities, discharge lamps, and hot-wire detectors for state preparation, evolution, and readout of atomic hyperfine states.  Over time, new techniques were developed, motivated by understanding of fundamental interatomic interactions and the interaction with light. Modern atomic sensors now utilize high-performance lasers, optics, and electronics to improve the sensitivity, in some cases reaching quantum limits.  They may also exploit quantum control of momentum states, a dramatic outgrowth of fundamental AMO research.  Future development of atomic sensors will continue to be intimately connected to both the maturation of the supporting technology tools and advances in fundamental understanding.

This Perspective illustrates the principles of operation of typical atomic quantum sensors, with an eye to the sensor development process and issues that arise in trade-offs between performance and practicality. We illustrate these principles through examples and highlight the supporting technology required for their implementation; this serves to illustrate the cooperation between state of art performance and the technological innovations that enable it.

%We elect to focus on representative examples and direct our readers to previously established literature and reviews for further details.  Our selections prioritize recent examples where possible.  Furthermore, we emphasize atom-based sensing as a natural outgrowth of the fundamental science generally punctuated in this journal. In fact, the atomic sensor advantage derives from detailed understanding of atomic interactions with light and the environment, topics that have dominated this journal for decades.

\subsection{Definition}\label{sec:def}
A broad definition of a ``quantum sensor" is ``a device, the measurement (sensing) capabilities of which are enabled by our ability to manipulate and read out its quantum states" \cite{Safronova2021QuantumElephants}. Atomic quantum sensors more specifically exploit the quantum nature of atoms (and light) to make measurements. Varying perspectives and classification criteria can be found in recent quantum sensor reviews focused on fundamental physics \cite{Safronova2021QuantumSensors,JacksonKimball2023ProbingSensors}, quantum information \cite{Degen2017}, biomedical applications \cite{Aslam2023QuantumApplications,Wu2022RecentApplications}, energy applications \cite{Crawford2021QuantumPerspective}, photonics-based sensing \cite{Pirandola2018AdvancesSensing,Kutas2022QuantumLight,Datta2025}, molecular-based sensing \cite{DeMille2024QuantumMolecules, Yu2021ASensing}, distributed sensing \cite{Zhang2021DistributedSensing}, and even particle physics \cite{Bass2024QuantumPhysics}.  
One theme among some of these references is an argument that the term ``quantum sensor" should be reserved for those sensors that exploit uniquely quantum properties such as entanglement and squeezing, lumped under an advanced quantum methodology or quantum-enhanced metrology. Such a narrow definition is not useful, as it would exclude almost all existing atomic clocks and magnetometers from consideration as quantum sensors.
The appropriateness of advanced quantum sensing is addressed in Sec.~\ref{sec:QE}. 

Our perspective is simple:  if the properties of a quantum system are being used to measure something else, then the system is a quantum sensor.  The quantum sensor may or may not have performance that is superior to a classical sensor in all aspects, but any noteworthy sensor must have some unique capability to distinguish itself and make it useful.  For a related and helpful discussion, see Safronova and Budker in Ref.~\cite{Safronova2021QuantumElephants}.

\begin{figure}
    \centering
    \includegraphics[width=1\linewidth]{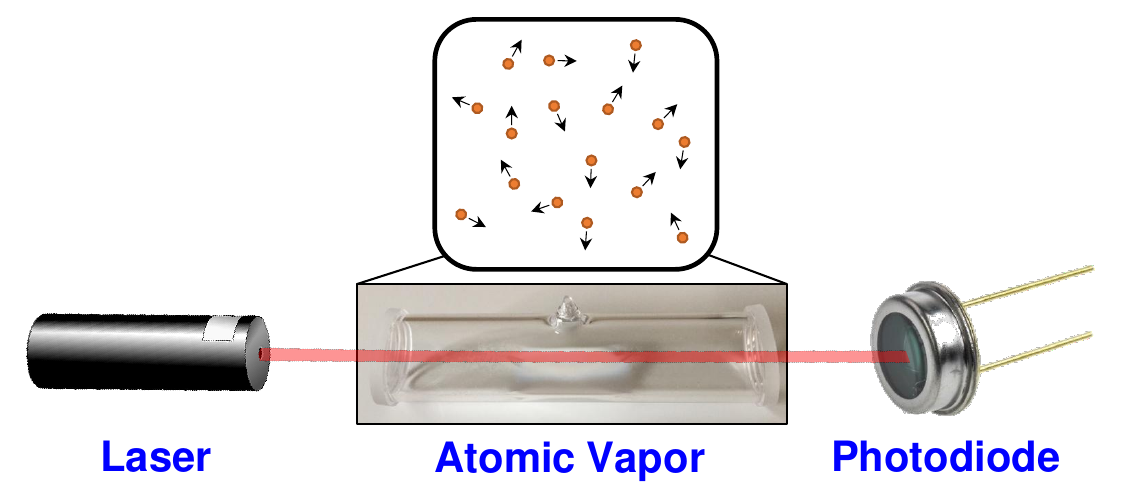}
    \caption{Representation of the essential elements of any atom-based sensor requiring a light source, vapor, and detector.}
    \label{fig:simple_sensor}
\end{figure}

The focus in this paper is on atom-based quantum sensors that are intended for use in a non-laboratory setting.  We also make a borderline tautological distinction between “quantum sensing” and a “quantum sensor”.  Most modern laboratory experiments, especially in AMO physics, use quantum sensing to address a scientific goal.  Such experiments are often of a remarkable degree of complexity and sophistication. 
For the purposes of this Perspective,  a quantum sensor is more specifically an apparatus that has been stripped to its bare essentials and simplified in order to make it compact and operable by a non-specialist, while retaining as much {\Edit of its performance }as possible\strike{ of its performance}.  Figure~\ref{fig:simple_sensor} illustrates a minimalist sensor containing (1) a light source, (2) a collection of identical atoms, and (3) a means of \strike{electromagnetic} detection{\Edit, usually optical}.  Such a quantum sensor lacks the large array of auxiliary measurement capabilities that a well-designed quantum sensing  laboratory apparatus has, but, if done right, can still exploit the remarkable properties of atoms to make ``useful" measurements.

%For the purposes of this Perspective, we view a quantum sensor is instead an apparatus that has been stripped to its bare essentials, and simplified in order to make it compact and operable by a non-specialist, while retaining as much as possible of its performance.  Such a quantum sensor rarely has the large array of auxiliary measurement capabilities that a well-designed laboratory apparatus utilizing quantum sensing has, but, if done right, can still exploit the remarkable properties of atoms to make useful measurements.

\subsection{Atom-Based Sensor Advantage}\label{sec:atom_adv}

\begin{figure}[htb]
    \includegraphics{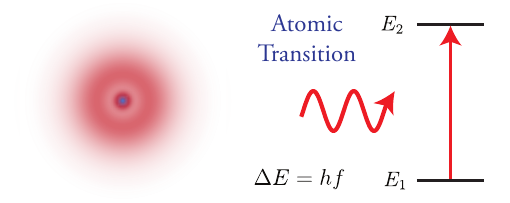}
    \caption{Depiction of \strike{an} {\Edit a rubidium} atom {\Edit showing outer shell 5$S_{1/2}$ (red) and inner shell (blue) electrons}.  {\Edit Representation of} \strike{with} photon interaction on {\Edit a} simplified 2-state energy level diagram.}
    \label{fig:atom}
\end{figure}

Vapor{\Edit -}phase atoms, manipulated and probed by light and other electromagnetic fields (Fig.~\ref{fig:atom}), constitute a versatile and powerful quantum \strike{system} {\Edit platform} for sensing.  The energy levels of most atomic states are known to 8 or more significant digits of absolute accuracy, and their dependence on external fields can also be predicted with high precision, extremely desirable properties for sensing.  Both conceptually and quantitatively, we may consider each atomic state of energy $E_n$ to be represented by a deBroglie clock ticking at a frequency $f_n=E_n/h$, where $h$ is Planck's constant {\Edit \cite{Feynman1985QED:Matter}}.  {\Edit Measurement outcomes depend only on energy differences, so} \strike{As always in physics, measurements} \strike{can only depend on energy differences, and} with rare exceptions, atomic sensors measure the Bohr frequencies $f_{nm}=f_n-f_m$ via interactions with light/fields of frequency $f\approx f_{nm}${\Edit ~\cite{Dirac1924NoteCondition}}.  For atoms in the vapor phase, these stable and predictable energy differences result in a high inherent precision and absolute accuracy for either stable reference standards (clocks) or well-defined shifts from environmental perturbations (electromagnetic and inertial sensors).  

%{\Lingering The Hamiltonian describing an individual atom along with its interaction with the environment defines a set of discrete energy levels that respond to internal or external stimuli.  A strategic choice of atomic system and energy level structure provides a strong basis for a clear determination of the quantity of interest.  Interaction with a photon at or near the transition provides a measurement of the frequency difference through $\Delta E=\hbar\omega$ and as such leads to all quantum sensors being connected to a frequency measurement of some sort (Fig.~\ref{fig:atom}).  Observation of the photon on a photodetector connects the quantum system to an electrical signal that can be interpreted.  Such sensors provide the most precise measurements with potential to achieve the best long-term stability as Nature has provided large collections of \strike{indistinguishable} {\Edit identical} atoms for repeatable measurements, and the atomic quantum state can be prepared and read using available modern laser and electo-optic tools.}

The most attractive feature of atom-based sensors is that they offer the opportunity to produce measurements with excellent sensitivity or long-term stability.  The qualifier ``opportunity" in the previous statement is worth emphasizing as the performance of a sensor depends heavily on how it is implemented (Sec.~\ref{sec:implementation}).  {\Edit Atoms not only have fixed, predictable properties, but also many atomic states have narrow linewidths, allowing for high sensitivity measurements.} \strike{Atoms and their constituent particles (electrons, protons, and neutrons) are guaranteed to have fixed, predictable properties anywhere, at any time.  This is accompanied by naturally narrow transition frequencies (in the absence of decoherence), where the large change in response near the resonance enables high sensitivity measurements.  Similarly,} {\Edit T}ransitions tied directly to fundamental atomic structure enable atom-based sensors to serve as standards that do not rely on external calibration for stability and accuracy.  

Put another way, atoms make excellent quantum sensors because they are: (a) \strike{indistinguishable} {\Edit identical}, (b) isolatable, (c) interface-able, and (d) intelligible.

\strike{{\it a. Indistinguishable}  The principle of indistinguishability of fundamental quantum particles is at the heart of quantum mechanics.Isolated atoms are guaranteed to have fixed, predictable properties at any point in spacetime.  A well-designed atomic sensor can therefore be insensitive to manufacturing variations of the primary sensing element, leading to a reproducibility that in principle  is directly traceable to the fundamental physical properties of atoms.  }

\paragraph{Identical} Atoms of the same isotope are identical.  Well-designed atomic sensors can therefore be insensitive to manufacturing variations of the primary sensing element, leading to a reproducibility that{\Edit ,} in principle{\Edit ,} is directly traceable to the fundamental{\Edit ,} \strike{physical} {\Edit immutable} properties of atoms.  

\paragraph{Isolatable} In order to realize the stability and accuracy of a sensor, atoms must be well-isolated from sources of systematic errors, usually arising from their interactions with the outside world.  Wall interactions represent one such connection where laser cooling and trapping can completely isolate the atoms from the walls.  Remarkably, even a warm vapor confined \strike{by} in a glass cell remains substantially insensitive to energy shifts from wall collisions (with a few exceptions).  Stray electromagnetic fields can be shielded to some extent, but their well-understood interactions often allow atomic sensors to reliably manage their systematic effects.  A famous example is the atomic Cs clock, in which the energy difference between the two clock energy levels is independent of magnetic field (to first order). Uniquely, atomic vapors allow direct quantum measurements of inertial quantities through manipulation of momentum states.

\paragraph{Interface-able}
The ability to remotely manipulate and interrogate atomic quantum states using light is a key enabling feature of most atomic sensors.  The effective area of an atom interacting with a resonant photon is of order $\lambda^2$, allowing atomic states to be manipulated and detected by light with exquisite sensitivity.  Technologies such as compact, tunable diode lasers, electro-optic modulators, and ultrasensitive detectors provide an array of powerful tools that allow practical atom detection and manipulation at or near quantum limits.

\paragraph{Intelligible} In most cases, atomic interactions can be realistically and quantitatively understood directly from first principles.  The Hamiltonians governing atoms and their interactions with electromagnetic fields are extremely well-characterized, and even many important decoherence mechanisms have been well-studied allowing their consequences to be predicted with a high degree of confidence.

%\Thad{ADD STATEMENT ABOUT Synthetic atoms (quantum dots, NVs) being outside the scope of the paper.  Is this the best place to put it?}

%{\Edit These properties distinguish isolated atoms from other classical or quantum systems.  Color centers (diamond nitrogen vacancy, silicon carbide etc.) or quantum dots may embody some of these traits, but only to a significantly lesser extent than isolated atoms.  In certain situations, they may be considered artificial atoms, and as such, are outside the scope of this work.  It is also worth noting that solid state/condensed matter systems made of atoms can also be well-understood from first-principles, but exhibit band structure with significantly wider transitions and are subject higher levels of interparticle interactions and impurities in lattice structure not present in isolated atoms.}

\subsection{Atomic Sensor Types}\label{sec:devices}

An ongoing promise and challenge of atomic sensor research and development is to realize more atomic sensor advantages in reproducible hardware, particularly such that it can be used by a non-expert outside the laboratory. While the finished sensor ma\strike{n}y incorporate a minimum number of elements as in Fig.~\ref{fig:simple_sensor}, embedded within a deceptively simple housing is the coherent manipulation and readout of quantum states in a thoughtful way to achieve a robust measurement of interest.  

\begin{table}[btp]
    \centering
    \begin{tabular}{ccccc}
        Property & Mechanism &$\partial S/\partial E$ & QPN\\ %& Reported In\\
        \hline
        \hline
        Frequency &  Microwave & $( hf)^{-1}$ & $1\times10^{-13}/\sqrt{\rm Hz}$\footnote{$f=6.8$~GHz ($^{87}$Rb $|1,0\rangle\rightarrow |2,0\rangle$), $N=10^6$ {\Edit atoms}, $T_2=50$~ms}\\
        $(\Delta f/f)$&  Optical & $( hf)^{-1}$ & $8\times 10^{-17}/\sqrt{\rm Hz}$\footnote{$f=385~$THz (Rb $5S\rightarrow5D$), $N=10^8$ {\Edit atoms}, $T_2=238~$ns}\\
        \hline
        B-Field & OPM &$\mu_B^{-1}$ & 10 aT/$\sqrt{{\rm Hz}}\footnote{$\mu_B=e\hbar/2mc,N=10^{14}$ atoms, $T_2$=10~ms}$\\ %& fT/$\sqrt{{\rm Hz}}$\\
        E-Field  & Rydberg &$d^{-1}$ &15 pV/cm/$\sqrt{\rm Hz}$\footnote{$d=n^2 ea_0$, $n=50$, $N=10^6$ cold atoms, $T_2=10~\mu$s}\\ %& $(\mu V/cm)/\sqrt{{\rm Hz}}$\\
        Acceleration & LPAI &$ (\hbar k_{\rm eff}T/4)^{-1}$ &  4~ng$/\sqrt{{\rm Hz}}$\footnotemark[5]
        
        \footnotetext[5]{$N=10^6$ cold Rb, $k_{\rm eff}=4\pi/\lambda$, $T_2=4 T =80$ ms, $v=$20 cm/s}\\
        Rotation & LPAI &$ (\hbar k_{\rm eff}v T/\sqrt{2})^{-1}$ & $0.0003^\circ/\sqrt{{\rm hr}}\footnotemark[5]%\footnote{Inertial navigation units of $^\circ/\sqrt{{\rm hr}}$ used instead of nHz/$\sqrt{{\rm Hz}}$}
        $\\
        \hline
    \end{tabular}
    \caption{Atomic sensor types, with order-of-magnitude estimates of quantum projection noise (QPN) \strike{floors }for cm-scale vapor packages, see Sec.~\ref{sec:SQL}. The exceedingly small fundamental limits represent the potential for high fidelity measurements.  In practice, technical noise (internal or environmental) is almost always orders of magnitude greater than these values.}
    \label{tab:devices}
\end{table} 

Examples of practical atomic sensor types \strike{that have been best developed to date} are listed in Table~\ref{tab:devices}.   Atomic clocks, whether microwave or optical, use direct measurement of an atomic Bohr frequency to produce a frequency reference that exhibits all the advantages of the atomic system discussed above.  As such, atomic clocks have long been the established frequency standard and best reference to timekeeping, even leading to the SI definition of the second.  Since the Bohr frequency is a fundamental internal property of the atom, clock designs seek to completely isolate the atoms from environmental interactions, which can leave relativistic effects as \strike{the} leading non-technical perturbations.  The highest precision clocks also directly observe changes in the geopotential.

The other four atomic sensor types exploit atomic sensitivity to electromagnetic ($\vec{E}$ and $\vec{B}$) and inertial ($\vec{a}$ and $\vec{\Omega}$) fields.  Here, ideal sensor designs intentionally expose atoms to the field of interest without introducing additional sensitivities to other interactions.  In addition, these measurements of vector quantities can either be {\Edit ``}clock-like{\Edit "} and measure the magnitude of the relevant field, or can serve as vector sensors to detect one or more components of the field and/or field gradients.  For example, optically pumped magnetometers (OPMs) utilize the Zeeman effect to produce and detect a coherent precession of atomic electrons about an external magnetic field.  Rydberg sensors exploit the large dipole moment $d\sim n^2ea_0$ of atomic Rydberg states to detect electric fields.  Light pulse atom interferometers (LPAIs) utilize matter wave interference of laser-cooled atoms to detect linear acceleration and/or rotation based on the interferometer geometry.

\section{Quantum Sensing with Atoms}\label{sec:atomSensors}
By definition, a quantum sensor is governed by and exploits the laws of quantum mechanics.  The Schr{\"o}dinger or wave approach to quantum mechanics is intrinsically a differential equation that takes an initial condition, propagates that initial condition through time, and gives a final result that can be measured.  Thus, any quantum sensor must provide a method for producing a low-entropy initial state, allow a controlled evolution (\strike{both} coherent and{\Edit /or} dissipative) of that state for some time, and finally make a measurement on that state that reveals the quantity of interest.

%\Thad{ add clarification about our model not reflecting some types of entangled sensors or collective state...}

\subsection{Quantum State Preparation}\label{sec:state_prep}
As atoms have more complicated energy level structure than the depiction in Fig.~\ref{fig:atom}, an atom-based sensor must prepare the atoms in  low-entropy initial states suited to the measurement.  In some cases, the goal is to produce zero-entropy pure states, such as a single Zeeman sublevel in the hyperfine manifold of the ground state, a coherent superposition of such states, or a well-defined linear momentum state or superposition.  In other sensors, one may be content (and the sensor performance even optimized) by simply producing population differences or incoherent superpositions of Zeeman sublevels, which subsequently evolve into coherences.

The specific varieties of quantum state preparation to reduce both the electronic and momentum degrees of freedom relevant to atoms are too many to be usefully summarized here; instead, several types are described.

\paragraph{Optical pumping}
Optical pumping \cite{Happer2010} is the use of light scattering to transfer order from light to the internal states of atoms.  Given the vagueness of this definition, it follows that the varieties of optical pumping configurations are essentially limitless.  Broadly speaking, in cold atoms where light scattering can be easily made to dominate over other forms of decoherence (collisions, magnetic field fluctuations, etc.) the atoms will tend to accumulate in ``dark" states that minimize the light scattering rate.  In hot atoms, optical pumping is not as fast due to Doppler and/or collisional broadening of the excited states, and usually results in a balance between light-induced ordering and collisional decoherence.

\paragraph{Velocity selection}
In some cases, thermal equilibrium naturally produces nearly perfect electronic-state population imbalances such that {\Edit electronic} state preparation {\it per se} may be unnecessary.  This is true in systems involving excited electronic states, such as two-photon optical clocks and Rydberg atom rf receivers.  Instead, a range of velocity classes are selected using the direction of light and the Doppler effect to invoke resonant conditions within a subset of the momentum distribution of the atoms. This is particularly important for Rydberg sensors \cite{Adams2020}. %\cite{Weichman2024}
Note that in a multi-photon process even if the net wavevector ${\bf k}={\bf k_1}+{\bf k_2}+\ldots$ is small so that the multiphoton resonance condition is effectively Doppler free, the process may yet be velocity selective if the intermediate state \strike{wavevectors} {\Edit Doppler shifts}, ${(\bf k_1}+{\bf k_2})\cdot v$ say, are \strike{substantial} {\Edit comparable to the intermediate state linewidths \cite{Ryabtsev2011Doppler-Transitions}}.  This is the case for the Rb 5$S-5P-5D$ 2-photon optical clock \cite{Maurice2020}, where the 5$P_{3/2}$ state lies very nearly half-way between the 5$S_{1/2}$ and 5$D_{5/2}$ states. %{\Edit , a narrow, Doppler-free transition can be driven with a modest amount of optical power due to the nearby-lying $5P_{3/2}$ state}.

\paragraph{Laser Cooling}
Laser cooling \cite{Schreck2021LaserGases} can be considered a generalization of optical pumping to the external or motional degrees of freedom of the atoms as entropy is removed from the atoms by light scattering or spontaneous emission.  Again, there are a large number of different schemes for laser cooling, the most effective of which utilize an interplay between coherent (AC Stark effect) and incoherent (scattering) state-dependent forces. Temperatures of order 10 $\mu K$ by sub-Doppler cooling are generally easily attained in almost any laser cooling apparatus on millisecond time scales.  More complex techniques such as Raman sideband cooling to temperatures \strike{approaching or exceeding} {\Edit below} the recoil limit are also possible.

Laser cooling can also be used as a starting point for the production of high phase-space density clouds using evaporative cooling from conservative laser or magnetic traps.  The evaporative cooling process is typically much slower than laser cooling, but is the best-known way to reach the lowest possible temperatures and highest densities at the cost of sensor bandwidth. 

\paragraph{Coherent Population Trapping}
A specialized form of optical pumping, Coherent Population Trapping (CPT, also known as Electromagnetically Induced Transparency, EIT) uses light fields, sometimes modulated, to drive the atoms into a steady-state which is a coherent superposition of the atomic eigenstates \cite{Finkelstein2023AVapor}.  In a common approach, the signature of the coherent superposition is an altering of the transparency of the atoms, giving a simple yet sensitive means of monitoring the formation of the atomic superposition.  The commercially available Chip-Scale Atomic Clock (CSAC), for example, utilizes CPT as the fundamental state preparation/detection mechanism for a compact hot atom microwave clock \cite{Lutwak2009}.

\subsection{Quantum State Evolution}\label{sec:evolution}
Following high-fidelity initialization, the quantum state evolves under a Hamiltonian containing all the interactions present.  Atoms are ``isolatable", allowing the sensor to be configured in a such a way that the quantity of interest is the dominant interaction; however, this isolation is rarely complete and a deep understanding of the remaining interactions with the environment remain key to understanding the operations and limitations of atomic sensors.  

A quantitative description is usually necessary and is provided through the quantum density matrix $\rho$, which obeys the master equation
\begin{equation}
    d_t\rho=-\frac{i}{\hbar}[H,\rho]+{\cal L}\rho,\label{eqn:densitymatrix}
\end{equation}
where $H$ represents the Hamiltonian of the ``isolated" atom.  This term is deterministic and usually includes the interaction that the sensor is trying to measure as well as terms describing the manipulation of the state evolution by external fields.

The remaining interactions with the outside world generally cause dissipation or decoherence, represented by the Liouvillian ${\cal L}$.  This can include optical pumping and/or cooling terms, plus effects introducing noise such as ambient field fluctuations, inevitable magnetic field gradients, and vibrations.  These interactions heavily influence the coherence time $T_2$ for the measurement, and may also produce systematic shifts.  For sensors involving excited states, ${\cal L}$ includes the atomic interactions with the vacuum modes of the electromagnetic field.

As atoms are also ``intelligible", it is a great advantage of atom-based quantum sensors that one can realistically and quantitatively understand $H$ and $\cal L$ from first principles, with a minimum reliance on phenomenological descriptions.

\subsubsection{Coherent Evolution}
Most atomic sensors use the well-understood interactions between atoms and electromagnetic fields to produce and measure coherences that evolve at a well-determined Bohr frequency.  A primary tool for manipulating quantum evolution of atoms includes electromagnetic pulses at frequencies ranging from audio to microwave to optical.  Of particular note are microwave and rf pulses used for Rabi manipulation of atomic Zeeman and hyperfine levels.  In many cases, it is convenient to replace such pulses by equivalent light pulses using stimulated Raman interactions.  

It is usually convenient to conceptually separate the Hamiltonian into $H=H_0+H_1$, where $H_0$ establishes the energy eigenstates, and $H_1$, often produced by oscillating fields, drives coherences.  If coherence has been established by appropriate pulses, free evolution ``in the dark" allows the state to evolve under the simplest possible, most well-controlled $H$, limiting systematic uncertainties from strong control fields.  This is in contrast to continuously operating sensors where the influence of control fields persists throughout the measurement time period, offering a simpler control sequence if $T_2$ is not significantly reduced.

Details are specific to each type of sensor.  To highlight the breadth of approaches, some examples of coherent evolution are:

\paragraph{Microwave clocks} The celebrated ``0-0" transition between the $m_F=0$ Zeeman levels of the two ground hyperfine states of the alkali-metal atoms represent a convenient transition for a clock \cite{Wynands2005AtomicClocks}.  The Bohr frequency for this case is quadratic in magnetic field for sub-gauss fields, rendering the clock frequency insensitive to magnetic field fluctuations.  The coherence between these ``clock states" can be driven by standing-wave or propagating microwave fields, or optically by stimulated Raman transitions driven by off-resonance laser light that is modulated at a sub-multiple of the clock frequency.

\paragraph{Two-photon optical clocks}  Counter-propagating optical fields \strike{to }drive a two-photon coherence between the ground $S$ state and an excited $S$ or $D$ state \cite{Martin2018}.  In this configuration, first-order Doppler effects are generally canceled by a counter-propagating geometry, allowing the use of hot atoms without substantial performance trade-offs. The 4-5 order of magnitude greater clock frequency gives a natural suppression of Zeeman and AC-Stark shifts at the same clock stability relative to microwave clocks.

\paragraph{Magnetometers} Coherent precession of the ground-state atomic spin in magnetic-sensitive states produces a frequency proportional to the magnetic field $\omega=\gamma B$ %{\Edit where $\gamma$ is the Larmor frequency }
and can be {\Edit observed} \strike{operated }in warm alkali vapors \cite{2013OpticalMagnetometry}.  The most sensitive {\Edit spin-exchange relaxation-free} (SERF) magnetometers operate typically within a few nT of zero field.  The fT- to pT-scale transverse fields being detected produce tiny coherences which are equivalent to a small rotation of the atomic angular momentum away from the direction of spin-polarization.  In larger fields, the spins precess many cycles before becoming dephased, allowing the precession frequency to be detected, proportional to the magnitude of the magnetic field.

\paragraph{Rydberg {\Edit atom} electric field receivers}  Multi-photon coherences between the ground state and a high-lying Rydberg state connect electric field variations to detectable ground state population changes that can be conveniently exploited in room temperature alkali vapors \cite{Adams2020}.  As the multi-photon processes are resonantly enhanced by nearby intermediate states, the process is velocity selective and excites a small fraction of the vapor into Rydberg states.  DC or resonant AC fields shift the Rydberg energy levels, indirectly altering the phases of the individual coherences.  To make the detectors phase sensitive, a local oscillator is used to provide a phase reference for the weak detected field.

\paragraph{Inertial sensors} Matter wave interferometers \cite{Bongs2019} use velocity-sensitive stimulated Raman or Rayleigh scattering  to transfer photon momentum to the atoms, placing the atoms in coherent superpositions of momentum states.  The pulses act as beamsplitters and mirrors to compare the \strike{the }accumulated Bohr phases along the various paths in an interferometer.  Relative populations in the output momentum states provide the acceleration and/or rotation of the apparatus relative to the freely-falling atoms up to multiples of $2\pi$.

%It is worth noting that there are many known ways to exploit the coherent evolution of the quantum state for sensor applications including some that have yet to be taken advantage of.  Further details on specific illustrative sensors are included in Sec.~\ref{sec:examples}.

\subsubsection{Decoherence}\label{sec:decoherence}
%The study of decoherence in quantum systems is a broad field in itself with many reviews extending into the philosophical \cite{Zurek2003,Suter2016Colloquium:Noise,Frowis2018MacroscopicImplementations,Schlosshauer2019}.  
The low-entropy initial state and its subsequent coherent evolution inevitably degrade as a result of experimental imperfections or other interactions with the environment.  For atomic sensors, decoherence mechanisms represent the key signal loss mechanism with the coherent evolution time $T_2$ setting the fundamental single-atom resolution of the sensor, $\Delta E\sim h/T_2$.  This presents a particular challenge for sensors as they may be exposed to the physical environment in a less than ideal way for the act of sensing.  A few common types of decoherence are described below.

%For an atomic vapor, key decoherence mechanisms can be grouped into (a) ballistic motion $\sigma\sim v_{th}/\lambda$, (b) diffusion $\sigma\sim D/l^2$, (c) transit-time broadening $\Gamma\sim v_{th}/\omega_0$ or $D/\omega_0^2$, and (d) collisional effects (notably spin-exchange and spin-destruction) as described in Ref.~\cite{Finkelstein2023AVapor}.  While a particular experiment will focus on specifics, our perspective highlights common limitations relevant to atomic systems and highlight some relevant examples below.  

\paragraph{Transit-time broadening} In the absence of collisions or other decoherence mechanisms, the time the atoms can be observed before leaving the observation volume limits measurement resolution. For cold atoms with $\sim$cm/s velocities, the transit time is typically set by gravity, the spatial extent of the interaction region, and the width of the velocity distribution.   If the atoms are velocity selected and released as is common in atomic fountain clocks and atom interferometers, the transit time can be further extended.  Another approach unique to atoms overcomes the limitations of gravity by retaining atoms in conservative traps without significant perturbations.   Great ingenuity has gone into devising methods such as magic wavelength traps where the AC Stark shifts of the relevant clock states are equal to enable long coherence times at a fixed location in space.  This approach has proven particularly fruitful for optical lattice clocks, albeit with substantial increases in complexity.  

For hot atoms with velocities $\sim$300~m/s, transit times are $\sim 30~\mu$s in a cm-scale evacuated vapor cell.  This generally limits the fundamental linewidths achievable in optical spectroscopy, in devices such as 2-photon optical clocks and Rydberg sensors.  

%DISCUSS For beam based experiments, the decoherence time can be limited by the time of flight if spatial uniformity of magnetic fields etc. can be addressed.  Ultimately, this comes down to the physical footprint of the beam apparatus and the temperature/velocity of the atom source.

\paragraph{Diffusion} For ground-state sensors, coherence times are generally a trade-off between reducing transit times by adding buffer gases, balanced against increased collisional relaxation.  It is remarkable that the cross section for spin-relaxation of alkali-metal atoms
with light, closed-shell atoms and molecules such as He, Ne, and N$_2$ allow for confinement with buffer gas pressures on the order of 1 bar, where velocity-changing collisions occur rapidly compared to natural lifetimes so that transit broadening is completely eliminated.  Under these conditions, the atomic motion is diffusive, with diffusion coefficients of order $D\sim$ 0.2 cm$^2$/s, and allows for spin-relaxation times of order 0.1~s in a 1 cm$^3$ volume depending on details of the pressure and buffer gas species.

An alternative approach to long spin-storage times in hot vapors is the use of anti-relaxation coatings \cite{WuRMP2021,Chi2020AdvancesCells}.   These specialized coatings enable many thousands of bounces of atoms off the cell walls before spin-relaxation occurs.  The coherence times can be thereby extended to approach $\sim 1$~s scales for cm-sized cells.
As the atoms average over the entire cell volume, a dramatic reduction in sensitivities to spatial field gradients (optical, magnetic etc.) occurs.

\paragraph{Field gradients} Non-uniformness, most commonly of magnetic and laser fields, is important for all types of atomic sensors, that lead to inhomogeneous broadening for stationary atoms and reducing coherence times for atoms moving through the gradients.  A recent comprehensive analysis of the systematic shifts and decoherence from diffusion through gradients is Ref.~\cite{Sheng2014NewComagnetometers}. Techniques such as spin-echoes can be useful for mitigating the effects of gradients \cite{Solaro2016CompetitionInterferometer}.

\paragraph{Collisions} One of the key driving motivations for the development of laser cooling was the suppression or near elimination of collisional decoherence mechanisms.  Indeed, with the exception of clocks operating at the highest levels of precision, cold atom sensors can typically be considered to be collision-free.  For hot atom sensors, the much greater atom numbers come with the need to cope with collisional effects.  Collisions can cause Bohr frequency shifts as well as dephasing and/or relaxation of the atomic coherences being generated in any sensor.

A long-studied example of the impact of collisions is on the species- and temperature-dependent frequency shifts from buffer gases on the $|F,0\rangle\rightarrow|F+1,0\rangle$ state in alkali atoms, relevant to microwave clocks \cite{Arditi1961,Ha2021}.  Study of these shifts has resulted in gas mixtures that can cancel leading order shifts at a given temperature. A persistent limit to the long-term clock stability remains finite drifts that still occur as cell conditions slowly change with time.

Spin-exchange collisions $A(\uparrow)+B(\downarrow)\rightarrow A(\downarrow)+B(\uparrow)$ represent an important decoherence mechanism in alkali-metal atoms and occur with a rate coefficient that is close to the gas-kinetic limit, $k_{se}\sim 10^{-9}$ cm$^3$/s. For example, hyperfine coherences $\langle\ket{F,m}\bra{F',m}\rangle$  are rapidly destroyed by spin-exchange collisions.  However, since spin-exchange collisions preserve the total spin of the colliding pair of atoms, the total spin of an atomic ensemble is not subject to spin-exchange decoherence \cite{Happer1977}.  In some sensors such as SERF magnetometry \cite{Allred2002}, collisions are embraced rather than avoided.  This is also true in a recent demonstration of light storage of 1~s in spin-polarized alkali vapor \cite{Katz2018}.

Spin-exchange collisions have a number of other remarkable features that can be exploited rather than avoided.  In the limit of rapid spin-exchange collisions, the density matrix is a spin-temperature distribution $\rho\propto\exp(\beta\cdot F)$  consistent with maximizing the  entropy of the angular momentum states.  Certain coherences, like  $\langle\ket{a,a}\bra{a,a-1}\rangle$ and  $\langle\ket{a,a}\bra{b,b}\rangle$, where $a=I+1/2$ and $b=I-1/2$, are not subject to spin-exchange decoherence in a fully polarized sample \cite{Jau2004}.  A recent detailed study of spin-exchange decoherence of Rabi flopping of hyperfine coherences revealed a variety of surprising dynamical features \cite{Kiehl2023}.  

%Rev Mod Phys Article on Decoherence \cite{Zurek2003}
%Review on Decoherence \cite{Schlosshauer2019}\\

\subsubsection{Simulation}\label{sec:sim-atom}
Atomic systems are ``intelligible"--the interaction of atoms with electromagnetic fields and collisions between alkali-metal atoms and common buffer gases have been extensively studied and are generally well-understood.  Realistic computer simulation of atomic sensors, while never a replacement for experimental investigation, can anticipate problems and guide understanding and interpretation. 

%Many sensor applications focus on a small number of number of energy levels interacting with resonant to near resonant interactions.  These can be modeled with a small number of equations using analytical methods or standard computational methods.  The qualitative insight into the sensor operation  builds intuition toward measurement optimization; however, the high sensitivity and high accuracy provided by atom-based sensing leads to more subtle dependencies than can easily be captured by these models leading to introduction of additional nearby levels or more complete simulation of the density matrix.

A classic example is understanding warm vapor atomic clocks based on coherent population trapping.  A\strike{n} treatment digestable at the undergraduate level can be produced by analyzing wavefunctions in a three state system \cite{Belcher2009AtomicLaboratories}.  This provides powerful insights and intuition, but it not capable of predicting important quantities such as linewidth and constrast.  Introducing an additional fourth ``trap" state to represent the collection of states indirectly involved can provide accurate predictions of these quantities \cite{Vanier2005AtomicReview}.  A more complete treatment using the density matrix provides the highest fidelity predictions and would be necessary to tease out subtle clock shifts and instabilities.

A variety of software packages are publicly available  for modeling of optical pumping and optical Bloch equations in the presence of essentially arbitrary low frequency fields.  In some cases, this has been shared publicly as documented software packages such as Rydiqule \footnote{{h}ttps://rydiqule.readthedocs.io/en/latest/}, ARC \footnote{{h}ttps://arc-alkali-rydberg-calculator.readthedocs.io/en/latest/}, A*D*M \footnote{{h}ttps://www.rochesterscientific.com/ADM/}, and {\it Optically Pumped Atoms} \cite{Happer2010}.  For an introductory tutorial on density matrix calculations using Python without black-box codes, see Ref.~\cite{Downes2023}.  An additional tutorial on laser spectroscopy and using available software packages is also available \cite{Pizzey2022}.

\subsection{Detection \& Readout \label{sec:det}}
An ideal quantum sensor has its technical noise contributions suppressed so that state readout performs at or near  quantum limits.  As the detection of atoms in modern sensors is almost always mediated through photons, quantum fluctuations of both atoms\strike{, quantum projection noise (}{\Edit --}QPN\strike{), }{\Edit --}and photons\strike{, }{\Edit --}photon shot noise (PSN)\strike{, }{\Edit --}contribute to the fundamental detection limit.  Based on the number of photons scattered $N_{ph}$ per detected atom, one of these two limits will usually dominate the observed noise floor if all other technical contributions are overcome. For instance, if $N_{ph}\gtrsim1$, QPN will dominate and if $N_{ph}\lesssim1$, then PSN will dominate.  These considerations usually hold whether the state readout is accomplished by photon scattering or by index of refraction measurements.

For cold alkali-metal atoms, shelving with fluorescence detection is a simple and effective means of high signal-to-noise (SNR) readout \cite{Dehmelt1975ProposedII, Nagourney1986}.  Resonant light tuned to the well-resolved transition $nS_{1/2}(F=I+1/2)\rightarrow nP_{3/2}(F=I+3/2)$ can scatter typically thousands of photons before a Raman process moves the atoms to the $nS_{1/2}(F=I-1/2)$ states.  The number of scattered photons can be enhanced even further with circularly polarized light and a bias magnetic field to optically pump the detected atoms into the $(F,m)=(I+1/2,I+1/2)$ state where the Raman processes are further suppressed.  Collection of the fluorescent light by a high numerical aperture lens and detection with a high quantum efficiency camera or photodiode allows for state detection at the QPN limit as long as care is taken to suppress backgrounds from laser scatter and laser intensity and frequency fluctuations.  After the shelving measurement, a second measurement period with a repumper allows for determination of the atom number. The ratio of the two signals is independent of the detection efficiency of the optical system.

Shelving is not possible for hot atoms due to Doppler and/or pressure broadening making the excited-state hyperfine structure insufficiently resolved and the Raman scattering less state-selective.  Nevertheless, optically thick samples can be measured non-destructively and reach the QPN. Let the atomic sample have $N$ atoms in a volume $L^3$, with an optical linewidth of $W$.  State detection of the atoms employs a probe laser of flux $\Phi$ tuned near an atomic resonance. The observable might be the absorption or index of refraction, or, for detection of a spin component, the circular dichroism or Faraday rotation of the polarization of the light.

The line-center optical thickness $OD_0$ of the sample determines the degree to which the probing can be nonperturbative.  This is a critical issue for sensors requiring continuous readout.  In order to absorb a small fraction $f$ of the probe light,  the probe detuning will be increased such that $\Delta/W\sim\sqrt{OD_0/4f}$, and signals based on the index of refraction will be larger than absorption signals by the same factor.  The photon scattering rate per atom is $r=f\Phi/N$, so the photon shot noise SNR becomes $(f \Delta/W)/\sqrt{\Phi t}=\sqrt{ OD_0 Nr t/4}$  in detection time $t$.  Comparison with a typical quantum projection noise  $\sqrt{N (\Gamma_2+r) t}$
shows that continuous readout can be made non-perturbative only if the atomic cloud is optically thick.  If $t$ is short compared to the bandwidth $\Gamma_2$ of the atomic response, the photon shot noise is degraded by $t\rightarrow \Gamma_2^2/t$ so the optical depth also sets the \strike{ transition frequency below} {\Edit averaging time above} which the noise is QPN{\Edit -limited} and \strike{above} {\Edit below} which is PSN{\Edit -limited}.

These considerations are relaxed if the probe light is multi-passed \cite{Li2011,Sheng2013,Kuzmich1998} with a Herriot-like cell or, potentially, a resonant optical cavity.  The effective optical depth is enhanced by the number of passes, \strike{increasing the frequency at which the QPN transitions to PSN.}   {\Edit decreasing the averaging time at which the PSN transitions to QPN.}

\subsection{Standard Quantum Limit}\label{sec:SQL}

%{\Justin I want to delete these two lines in favor of leading with the following paragraph.} {\Lingering As all atomic sensors are spectroscopic measurements of the Bohr frequencies of the atoms, the energy difference is the underlying physical observable.  That the measurement process itself disrupts the coherent evolution of quantum systems even in the absence of other forms of decoherence reflects another fundamental property of quantum mechanics.}

As all atomic sensors are spectroscopic measurements of the Bohr frequencies of the atoms, the energy difference is the underlying physical observable.  For a single measurement on a single atom, the uncertainty principle gives an energy uncertainty $\Delta E\sim \hbar/T_2$. The coherence time $T_2$ depends on context, but can generally be considered the time the atom evolves before measurements or collisions reset its deBroglie phases. Repeating the measurement $M=\tau/T_2$ times over measurement interval $\tau$ for $N$ atoms reduces the uncertainty to $\Delta E\sim \hbar (T_2\sqrt{M N})^{-1}=\hbar(N T_2 \tau)^{-1/2}$.  For a quantity $S$ being measured that depends on the energy, the spectral density of the measurement is
\begin{equation}
\delta S\sim\frac{\partial S}{\partial E}\frac{\hbar}{\sqrt{N T_2}}   \label{eqn:one}
\end{equation}
where the units have the dimensions of ``$S$"/$\sqrt{{\rm Hz}}$. This expression represents the so-called ``standard quantum limit" (SQL){\Edit , expressed as a spectral density rather than an uncertainty}.   As coined in Ref.~\cite{Budker2023} for magnetometers, this expression represents ``Equation One" for quantum sensing.  According to Eqn.~\ref{eqn:one}, a million atoms with a 10~ms coherence time has a SQL of $6\times 10^{-18}$ eV in a 1 second measurement.  Such a fine fundamental energy resolution is an extremely attractive feature of atomic quantum sensors. 

The SQL serves as a representative benchmark for the statistical potential of a quantum sensor. The relationship between the SQL and systematic errors is of critical importance (Sec.~\ref{sec:sensitivity}). Technical noise and systematics have to be addressed in order to reach QPN limits in any practical sensor. 

%\Thad{DELETE?} {\Justin Consents} Equation~\ref{eqn:one} represents a sliding scale for the SQL based on the physical conditions imposed by $N$ and $T_2$.  Based on number of atoms and coherence time, the quantum limit for a measurement can in principle be adjusted.  This could lead one to conclude that measurements of exquisite sensitivity can be achieved merely by increasing these values; however, this ignores important technical noise sources and systematic effects.  By the same token, it is possible create a scenario for a ``quantum-limited measurement" by operating with artificially small $N$ and $T_2$.  This has been an opportune {\it modus operandi} for many fundamental studies of the quantum limit and approaches for squeezing and other advanced quantum measurements, but less optimum conditions for improved sensing (Sec.~\ref{sec:QE}).  

Table~\ref{tab:devices} lists SQL estimates of what might be possible for ``practical" devices limited to cases where the ``sensor head" is at the cm-scale.
A mix of warm and cold atom approaches are included. In most cases, it is an open challenge to meet these impressive performance levels in practical devices. Research on low-noise state preparation, coherent evolution, decoherence mechanisms, readout methods, and their practical implementation (the topic of Sec.~\ref{sec:implementation}) is critical to meeting this challenge.

%Certainly, more impressive values are possible if this constraint were lifted, as would be involved in laboratory-scale science measurements; however, these estimates, as presented, are already exceedingly small. In fact, it is this feature that contributes to atoms serving as an attractive quantum sensing platform, that they offer the potential to produce measurements with exquisite sensitivity before fundamental limits are reached.  Operating with such precision requires a detailed understanding of the fundamentals, often in understanding collisions, one of the predominant contributors to $T_2$.

\section{Atomic Sensor Implementation}\label{sec:implementation}
Realizing the ``isolatable" and ``interface-able" advantages of atomic quantum sensors in hardware is a key challenge, even in relatively benign laboratory environments. A practical quantum sensor must also operate in uncontrolled environments with constraints on size, weight, and power (SWaP){\Edit~\footnote{While SWaP is commonly used, the space community prefers to reference an object's mass over its weight, leading to the size, mass, and power (SMaP) acronym.} }while simultaneously addressing other sensor characteristics (bandwidth, dynamic range etc.). Thus, there is an obvious intimate connection between sensor implementation and its supporting technology.  How the simple conceptual structure of Fig.~\ref{fig:simple_sensor} is implemented in hardware is discussed below.  The details will differ from sensor to sensor, but there are general themes that are worthy of consideration, including that 
%It is worth noting that while the atoms remain static, 
the tools available to implement a sensor are always evolving, the topic of Sec.~\ref{sec:support_tech}.

\subsection{Complexity Cost}\label{sec:complexity}
Physicists and engineers have a demonstrated capacity to manage complexity, so there is a natural temptation to design practical quantum sensors with multiple feedback loops, adjustment knobs, and exotic, high-performance materials; however, measurement robustness correlates with measurement simplicity.  A ``complexity cost" is associated with the sensor design.  For example, electro-optic tools such as  spatial light modulators or digital mirror arrays 
offer the ability to programmatically manipulate spatial patterns of light and create painted potentials \cite{Henderson2009,Amico2021}, but each additional knob introduces additional environmental sensitivity, detrimental to noise and systematic effects, particularly in a dynamic environment.  Other benchtop staples such as liquid crystal waveplates and external cavity diode lasers produce similar concerns.

On the other hand, measurement complexity is necessary when it provides an essential enhancement to the sensor.  A practical sensor should be ``as simple as possible but no simpler" \cite{Calaprice2019TheEinstein}.

%This section highlights a few unifying themes relevant to physically realizing a practical sensor with existing tools. 

%Based on the underlying physics, platform influences not only SWaP, but also other sensor characteristics, such as sensitivity, bandwidth, dynamic range etc. depending on the measurement.  

%The physical realization of the anticipated quantum sensor performance represents the greatest challenge of atomic sensors.  The predictability of the atomic system can lead to well-defined set of physical conditions that support the prescribed measurement, but even in a relatively benign laboratory environment, these conditions require care and a detailed understanding of the hardware to achieve.  Our ``practical" quantum sensors must also operate in uncontrolled environments with constraints on SWaP, leading to additional nuances in the construction.  We leave general sensor principles to the examples in Sec.~\ref{sec:examples} and details to the broader literature.  Here, we highlight a few unifying themes in the hardware relevant to physically realizing a practical sensor.

\subsection{Species Selection}\label{sec:species}
Most atomic sensors utilize the heavy alkali atoms, primarily rubidium or cesium.  These atoms are attractive  for several reasons. First, they have simple, hydrogen-like energy level structures, and there is an extensive literature detailing almost any property involving light and collisional interactions of these atoms. 
Second, the availability of narrow-linewidth semiconductor diode lasers on the relevant $nS_{1/2}\rightarrow nP_{1/2}$ D1 and $nS_{1/2}\rightarrow nP_{3/2}$ D2 transitions in the near infrared is extremely convenient.  
Third, they can be easily produced and controlled in the vapor phase.    This includes high densities ($10^{12}-10^{14}/{\rm cm}^3$) at 80-130$^\circ$C ideal for vapor cells or moderate densities ($10^8-10^{10}/{\rm cm}^3$) closer to room temperature that are compatible with ultra-high vacuum for atom trapping.  Finally, hyperfine splittings of $3-10$ GHz are compatible  with high performance, inexpensive electronics.  

There are sometimes compelling reasons to use other atoms.   Recently, the remarkable properties of the  $^3P$ states of alkaline earth and other two-electron atoms have been exploited, which present narrow intercombination lines $^1S\rightarrow {^3}P$ attractive for next-generation optical atomic clocks.  Some ions offer similar simplified energy level structures with alkali-like electronic states in Yb and Group II \& IIB elements and alkaline-earth-like ions in Lu and Group IIIB elements \footnote{{h}ttps://iontrap.duke.edu/resources/ion-periodic-table/}.  Optical manipulation of these species requires lasers from the ultraviolet to infrared, increasing the complexity cost relative to the heavy alkalis, but laser technology advances continue to reduce this problem.
%The myriad of transitions across the are harder to access than the heavy alkalis, though solid state physics advancements are producing more narrow-linewidth diodes at relevant wavelengths.  A top candidate ion species is less clear cut.

\subsection{Platform Ramifications}\label{sec:platform}

There are four basic atomic sensor platforms illustrated in Fig.~\ref{fig:platforms} arranged with increasing SWaP: 
warm vapors, atomic beams, and trapped atom systems (neutral or charged).  The main features of these platforms are:

\begin{figure}[htb]
    \includegraphics{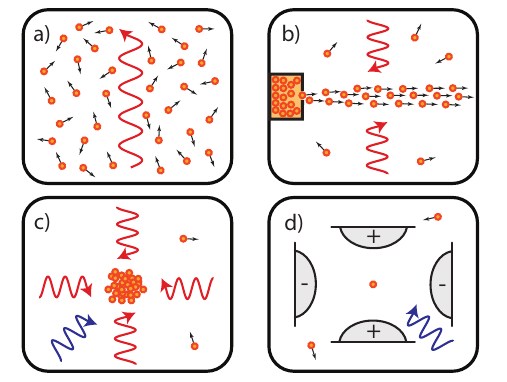}
    \caption{Atomic vapor sensing platforms: a) warm vapor cell of $N\sim10^{10}-10^{14}/{\rm cm}^3$ atoms with 3-dimensional rapid thermal motion and collisions b) an atomic beam geometry approximates one-dimensional motion with a modest number of background atoms c) neutral cold atom trapping isolates $N\sim10^7-10^9$ atoms in an UHV environment d) oscillating electric fields trap $N\sim10^0-10^1$ ions in an UHV environment.}
    \label{fig:platforms}
\end{figure}

\paragraph{Warm vapor} Sealed vapor cells, usually made of glass, may be heated in an oven to produce an equilibrium vapor density\strike{,} {\Edit (}$10^{10}-10^{14}/{\rm cm}^3${\Edit )}\strike{,} of alkali-metal atoms (Fig.~\ref{fig:platforms}a).  The sensor volume is dominated by the cell and associated oven as well as the optical path for delivering the lasers.

Evacuated cells, used for excited-state applications such 2-photon optical clocks and Rydberg electric field sensors, are transit-time limited to tens of microseconds and require sophisticated laser stabilization to achieve narrow linewidths for the multiphoton processes.

For magnetometers and some types of clocks utilizing the ground-state atomic hyperfine levels, the cells may be coated with anti-relaxation materials, or also contain an inert buffer gas mixture with favorable collisional properties in order to produce a long coherence time, typically tens of milliseconds. {\Edit Doppler- or pressure-broadened} optical linewidths in the GHz range imply that laser stabilization requirements are modest.  

\paragraph{Atomic beams}  Atomic velocities can be aligned along a common direction by collimating atoms from a heated reservoir or other funneling mechanisms to form an atomic beam (Fig.~\ref{fig:platforms}b).   A high vacuum (HV) environment ($<10^{-4}$~Torr) extends the mean free path beyond the length of the beam. 
%while the reservoir may maintain a higher pressure ($\sim 10^{-2}$~Torr).

Laser or microwave radiation applied transverse to the beam axis reduces Doppler-broadening from the axial velocity to the smaller transverse velocity profile and can be applied at various positions.  Fourier-limited ground state hyperfine transitions can be observed for use in microwave clocks, though optical linewidths are set by the excited state lifetimes, requiring more stringent laser stabilization and control relative to vapor cells.  Field homogeneity requirements are critical in the atomic interaction region which represents a small fraction of the volume.  A typical system is dominated by the vacuum enclosure (atom source, pump etc.) and  frequency-stabilized lasers.

These techniques can be applied to nearly any species and have been exploited in the cesium beam clock \cite{Bauch2003}.  An excellent review of thermal beam methods is available in Ref.~\cite{Ramsey1996}.

\paragraph{Trapped atoms}  Laser cooling and trapping of neutral atoms, as most simply implemented, replaces the atomic beam oven with a getter source or its equivalent, maintaining an ultra-high vacuum (UHV) from which $10^7-10^9$ atoms are laser cooled and trapped in the intersection region of several laser beams (Fig.~\ref{fig:platforms}c). % Cooling of atoms produces strong hyperfine energy shifts and can limit spectroscopic measurements.  
To avoid unwanted light shifts, a compact cloud of atoms cooled to a few $\mu K$ is usually released or moved to a conservative trap for sensing.  Detection is often destructive, so atoms are reloaded to repeat the process.  The complexity and sophistication of the required lasers for cooling, state preparation, and detection are greater than for warm vapor sensors.  The long coherence times and greatly reduced collisional effects, plus the access to manipulation of atomic momentum states, justify the complexity for high-precision clocks and inertial sensors.

Oscillating electric fields can collect small numbers of ions in Paul or Penning traps.  Modern chip-scale ion traps can be printed on planar surfaces with electrodes in custom configurations and operate within an UHV environment.  Several frequencies of light across the UV and visible spectrum are required to cool and manipulate the ion, nominally requiring only milliwatts of power, but with  $<$MHz  frequency stabilization requirements.  Ion traps are extremely useful for high-performance clocks.  The small atom numbers $N\sim10^0-10^1$ and high detection efficiency using shelving make QPN dominant.  Ion traps are also being explored for gyroscopes \cite{Campbell2017}.  

%\paragraph{Atomic beam} An atomic beam offers greater control over motion of the atoms in exchange for complexity in the instrumentation (Fig.~\ref{fig:platforms}b).  Here, a reservoir produces a flux $dN/dt$ of atoms through an aperture that approximates one-dimensional motion with minimal divergence into the transverse dimensions.  In a beam configuration, collisions are reduced and light interacting transverse to the beam mitigates Doppler effects where $T_2$ can be determined by transit times defined by beam velocity, beam diameter, and apparatus length.  

%\paragraph{Trapped atoms} Laser cooling and trapping reduces the velocity of a small number of atoms to cm/s velocities to a regime where Doppler shifts and XX are minimal.  

%motion in exchange for even more complexity in the apparatus.  Cooling atoms to Doppler and sub-Doppler temperatures enables observation of natural linewidths in three dimensions without Doppler-free spectroscopy techniques and allows dramatic reduction in collisional rates due to the reduced kinetic energy.  The supporting vacuum and laser stabilization infrastructure can isolate $N\sim10^7-10^9$ neutral atoms (Fig.~\ref{fig:platforms}c) or  $N\sim10^0-10^1$ ions (Fig.~\ref{fig:platforms}d) from the environment.

%\paragraph{Discussion}
%{\Justin 1/15/25 Justin Deputy - Make nice words about beams and ion traps}

In general, when the measurement involves only internal degrees of freedom of the atoms and the consequences of atom-atom interactions can be managed, the SWaP afforded by the vapor cell provides a clear advantage.  Examples include magnetometers and modest performance microwave clocks.  There are other instances where vapor cells cannot perform the measurement well, as a consequence of short transit times or mean-free paths, or systematic errors or dephasing from collisional frequency shifts.   Accelerometers based on matter wave interferometry benefit from slow atoms and a collision-free environment %\footnote{A unique vapor cell accelerometer has been demonstrated in Ref.~\protect\cite{Biedermann2017}, however, the velocity-changing collisions detrimentally impact its sensitivity below an interesting level.}.  
Likewise, for the highest precision microwave clocks, cold atoms are essential. 

%Cold atom systems control velocity states to greatly simplify the atom-atom interactions and are able to perform precise measurements at the expense of managing technology elements divorced from the measurement (ultra-high vacuum, multiple sub-MHz frequency-stabilized lasers etc.).  However, these systems access unique measurement regimes such as $N_{ph}\gtrsim 1$ such as in trapped ion systems allowing QPN to dominate over PSN in detection (Sec.~\ref{sec:det}).  Another example includes {\Justin Thad, how about one more favorite example?}

There are a small set of cases where both vapor cells are impractical and laser cooling is difficult.  Here, an atomic beam offers an alternative.  Examples include optical clocks based on the Ca intercombination line \cite{Fox2012,Shang2017CaClock,Tobias2025ColdSpectroscopy} where the reasonable vapor densities require extreme temperatures ($\sim 600^\circ$C) and calcium MOTs suffer from low peak optical densities and short trap lifetimes \cite{Mills2017}.  In other cases, alkali atom beams may overcome long-term systematics persistent in chip scale vapor clocks \cite{Martinez2023} or extend bandwidths in an atom interferometer to $\sim 1~$kHz \cite{Kwolek2022}, faster than cold atoms can be trapped to date.  

%  For inertial sensors and many clocks that cannot tolerate collisions, the additional complexity of cold atoms with more stringent technology requirements presents the only choice.

%{\Justin Better for hardware} Here, a heater and lasers used directly for state preparation and detection serve as the supporting technology whereas a cold atom system requires multiple elements divorced from the measurement process such as ultra-high vacuum as well as stabilized lasers to an auxiliary reference to support atom trapping. 

%\Thad{Sr vapor cell \cite{Pate2023}}

\subsection{Pulsed vs. Continuous Measurements}
The classic Rabi-Ramsey sequence \cite{Ramsey1990} that dominates the design of most precision measurements in AMO physics separates the state preparation, evolution, and readout functions in time as an inherently pulsed measurement.  In this case, the lasers used for preparation and readout are off during the evolution phase, suppressing interactions from the measurement process that could cause decoherence where in principle the precision of the measurement can be limited only by fundamental decoherence processes in the experimental environment.  Time separation allows state preparation to set atom-laser interactions as strong as possible, thereby driving the atoms into the desired initial state with maximum fidelity.  Such operations needed to prepare the initial state usually produce dissipation (laser cooling, optical pumping etc.), deleterious to the subsequent coherent evolution of the atoms.  Finally, the optical interrogation for best state detection reflects a complementary process to state preparation, leading to increases in decoherence, so pulsed measurements allow the flexibility to optimize state preparation and detection parameters in their respective time step.

A disadvantage of pulsed sensors is that the sensor bandwidth is limited to the Nyquist limit of $1/2T_c$, with $T_c$ being the repetition period of the experimental cycle.  Not only is the bandwidth limited, but also fluctuations above the Nyquist frequency are aliased into the detection band of the sensor, raising the noise floor at detection frequencies.  State preparation and readout intervals lead to dead-time in the measurement.  A partial mitigation strategy at the system level integrates multiple sensors as an array whose individual detectors have staggered pulse sequences, but this does not eliminate the problem of aliasing.  For laboratory measurements, such problems can be minimized with judicious selection of $T_c$ or filtering, but sensors operating in noisy environments may not have this luxury.

While nearly all cold atom sensors are pulsed due to short coherence times during the laser cooling process, for hot atoms there is the option to implement a structure where state preparation, evolution, and readout occur simultaneously and continuously, mitigating the dead-time problems.  Not surprisingly, continuously operated atomic sensors are in many respects complementary to pulsed sensors, though the considerations driving sensor design can be quite different.  For the greatest simplicity, both state preparation and detection may be implemented with a single laser.  This is exemplified by sensors using EIT/CPT.  For better performance and increased design flexibility, it may be advantageous to use separate lasers for state preparation and readout without paying too high a price in experimental complexity.  Whatever choices are made regarding state preparation and readout, they necessarily increase the fundamental decoherence rates and so a balance has to be  struck between high fidelity state preparation and readout versus long coherence times.

\subsection{Additional Hardware Considerations}
%Important  hardware considerations relevant to the  (a) laser, (b) vapor control, and (c) detector elements (Fig.~\ref{fig:simple_sensor}) are discussed.  A self-contained sensor also requires (d) an enclosed sensor package and (e) a controller with a user interface.

\paragraph{Optical Generation \& Manipulation} Semiconductor diode lasers represent the near universal choice for optical frequency generation in modern quantum sensors \footnote{Frequency-doubled fiber lasers are also leveraged to some extent}.  Sources are available at a variety of relevant atomic wavelengths and offer GHz-scale tuning by adjusting current and temperature, but must be stabilized or referenced to an atomic transition.  Laser cavities are susceptible to optical feedback, often requiring optical isolators.  A variety of optics and modulators are often necessary for controlling the spatial profile, polarization, and frequency of the light, as well as other functions such as frequency/phase/amplitude modulation, high-extinction shuttering, frequency shifting, and optical amplification.  Optical paths from the source to the delivery to the atoms must be managed and are often a significant contributor to sensor volume.  Improvements in {\Edit the volume required for} optical manipulation {\Edit can be achieved} using chip-scale photonic integrated circuits (PICs), see Sec.~\ref{sec:support_tech}. 

%Optical paths from the source to the delivery of the atoms must be managed using a variety of optics and modulators to adjust the spatial profile, polarization, and frequency as well as other functions such as frequency/phase/amplitude modulation, high-extinction shuttering, frequency shifting, and optical amplification as needed.

%\Thad{THIS SEEMS TO BE LACKING A DISCUSSION OF THE NECESSARY OPTICAL TRAINS.  Also, should fiber lasers be included?}

%Distributed feedback (DFB) lasers and distributed bragg reflector (DBR) offer $\sim80-200$~mW of power with $\sim$MHz linewidths while vertical cavity surface-emitting lasers (VCSELs) offer excellent quantum efficiency for low power applications, but are limited to outputs of several mW and linewidths of $\sim50$~MHz.  

\paragraph{Vapor Regulation}  The atomic vapor represents the primary sensing element and follows the types of implementations in  Fig.~\ref{fig:platforms}.  In all cases, the vapor is contained in an sealed vessel to isolate interactions with atmospheric gases, particularly reactive molecules such as oxygen and water.  Various glasses (with or without anti-reflection coatings) are favored for optical access, but metals or ceramics may be incorporated as a frame in an optical configuration to support the measurement.  Selection of low out-gassing and helium impermeable materials limit changes in the interior atmosphere over time.  Preparation involves heat and evacuation under vacuum to clean the interior surfaces and pump away contaminant molecules.  A source of atoms may be deposited following bake-out or released later from a chemical compound or other active source.  In some cases, inert buffer gases (N$_2$, He, Ar etc.), anti-relaxation coatings, getters, or active pumps are also included.  A final seal leaves a pinch-off; in some cases stemless seals are possible.  This volume commonly incorporates some form of thermal regulation to maintain equilibrium vapor pressures.

\paragraph{Detection} The prevalence of atomic transitions in the visible and near infrared wavelengths leads to the prominence of a silicon photodiode as the detector of choice favoring its compactness, low room temperature thermal current, and responsivity of 0.5 A/W ($\sim75\%$~quantum efficiency).  Faithful conveyance of the signal into a interpretable format represents the primary goal, starting with signal photon collection and scatter rejection and extending to electrical signal generation.  The photon to electron conversion is inherently an analog process, requiring a front end containing a highly linear transimpedance circuit.  Following the buffered output, the signal will be digitized with high-quality analog to digital conversion (ADC).  A design matched to the measurement \footnote{For instance, the detector white noise floor contributes over its full bandwidth, so a detector bandwidth $\gg$ measurement bandwidth leads to additional rms noise contributions in the measurement.} considers gain, bandwidth, and noise, though improvements in two parameters inherently limits the third.  Resources for circuit topologies and nuances in design can be found in Refs.~\cite{Graeme1996, Hobbs2009}.

%Internal amplification is available in  avalanche photodiodes (APDs) and photomultiplier tubes (PMTs) providing gains up to 100 to $10^7$ respectively with sizeable dark currents.  High voltages APDs are sensitive to overvoltage, PMTs sensitive to shock - less attractive for portable sensors, high voltage, fragility

\paragraph{Package/Enclosure} Unlike the laboratory breadboard system, a self-contained sensor requires a package \footnote{Using common jargon, ``physics package${"}$ refers to the container for the atomic vapor.  As the physics of the measurement requires the light generation, interaction, and readout, here we consider the entire sensor unit to be the package.} including an enclosure to limit unwanted interactions with the environment.  This can include electromagnetic shielding, hermetic seals, and internal thermal regulation.  Material selections and design choices provide a rigid internal support structure to maintain optical alignments from dynamic (shock, vibration, acceleration etc.) and thermal changes \footnote{A practical advantage occurs when thermal regulation is set above the environmental temperature range.  Stability is maintained at all environmental temperatures with feedback engaged and unnecessary power losses are avoided as heating requires less energy than cooling.}.  Internal interactions within the package must also be mitigated, in particular, magnetic gradients that can be introduced by magnetic materials found in most commercial electronics (nickel, iron, or Kovar) or current carrying wires.  A cable with electric and optical connections can be used to spatially separate a sensitive sensor head from the rest of the supporting hardware.  The package also manages how the sensor responds when disconnected from a power source, which can be a challenge for applications with ultra-high vacuum that rely on active pumps.

\paragraph{Controller} An embedded controller with an interface enables an operator to perform the measurement where internal technical manipulations are divorced from sensor use.  The controller manages startup, measurement, or data logging processes as well as any internal error handling.  The controller may manage auxiliary circuits for driving inductive loads, engagement of feedback loops, arbitrary waveform computation, or direct digital synthesis for DC to microwave frequencies.  Programmable changes are available through the interface, but a custom, low-level implementation leads to robust operation with minimal power consumption.  

\subsection{Sensitivity \& Accuracy Considerations}\label{sec:sensitivity}

While atomic quantum sensors generally have excellent potential for impressively low statistical noise, this is only really relevant if the sensor is sufficiently insensitive to sources of systematic errors.  Bias refers to the tendency of the sensor to give an incorrect value of the quantity being measured, in other words, a precisely reported but incorrect value.  The source of the systematic error may itself fluctuate, resulting in additional technical noise that also limits the statistical precision of the measurement.  

The sources of statistical and systematic errors can vary wildly across different types of sensors. Clocks and inertial sensors are primarily focused on long-term stability.  Electromagnetic sensors, in contrast, are usually seeking to detect transient phenomena and therefore care less about long-term drifts. Common sources of relevance to all atomic sensors include: magnetic field fluctuations and gradients, laser frequency and intensity noise, acoustic noise causing relative motion of optics and detectors, and electrical noise.  If the first order of business in an atomic sensor is to arrange for high sensitivity to the quantity of interest, a close second in importance is to design the sensor to suppress sensitivity to sources of systematic errors.  %and second of all to use sound low noise design wherever practical.

The performance of clocks and inertial sensors is commonly characterized by the Allan deviation $\sigma(\tau)$, a concept originally developed for oscillators but of general applicability to atomic sensors of various types.  Noise sources are typically characterized by different dependencies of the noise on frequency.  The Allan deviation maps distinct power-law frequency dependencies to different dependencies on averaging time.

\begin{figure}
    \centering
    \includegraphics[width=1\linewidth]{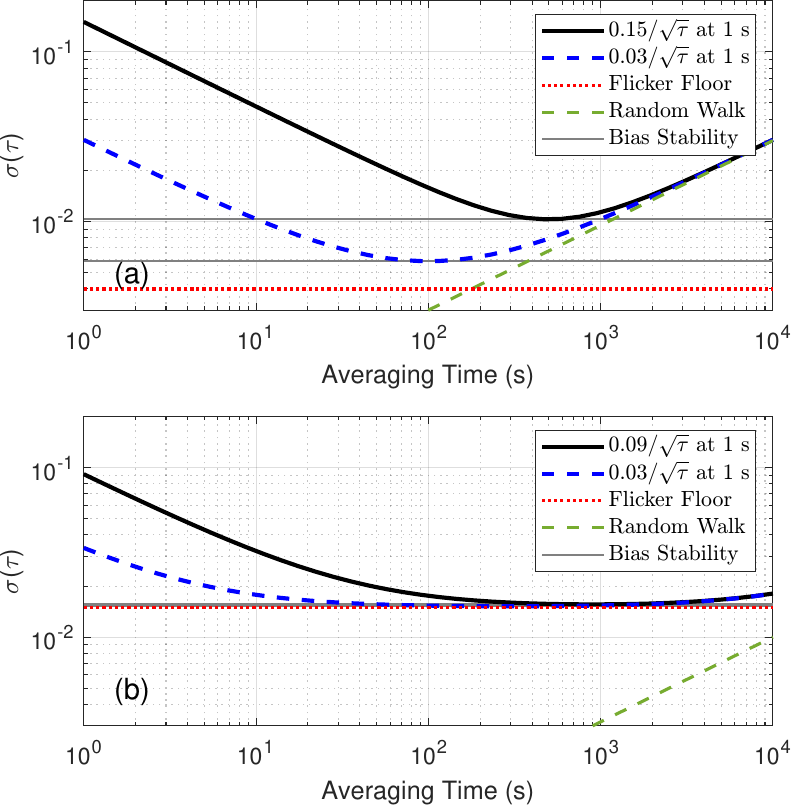}
    \caption{Allan deviations for hypothetical scenarios with different statistical {\Edit (white)} noise competing with {\Edit the same} technical noise to reach a bias stability at $\sigma_{bs}=\sqrt{2AB+C^2}$ over time $\tau_{min}=A/B$. (a) For small values of flicker floor, improved white noise improves bias stability.(b) Performance limited by flicker floor noise; in this case improved white noise offers only marginal improvement in measurement resolution, but reduces averaging time to the noise floor.}
    \label{fig:allan_deviation}
\end{figure}

It is informative to consider how white frequency noise interacts with systematic fluctuations that often scale inversely with frequency.  As a function of averaging time $\tau$, white noise scales as $\sigma=A/\sqrt{\tau}$ while the \strike{$1/f$} {\Edit random walk} noise {\Edit ($1/f^2$ noise)} obeys $\sigma=B\sqrt{\tau}$ and flicker noise {\Edit ($1/f$ noise)} follows a time-independent $\sigma=C$ {\Edit where $A$, $B$, and $C$ are coefficients that depend on experimental conditions}.  In the presence of these effects, the Allan deviation minimizes at an averaging time $\tau_{\rm min}=A/B$. Averaging for times longer than $\tau_{\rm min}$ degrades rather than improves the precision.  

The minimum value of the Allan deviation is known as the bias stability which in this case is $\sigma_{bs}=\sqrt{2AB+C^2}$.  The value of the bias stability, the ultimate resolution of the sensor, represents one of the key parameters characterizing long-term sensor performance.  As can be seen mathematically and pictorially (Fig.~\ref{fig:allan_deviation}), the bias stability is a balance between the white frequency noise and the competing technical noise sources.   The linear dependence in $\tau_{min}$ and square root dependence on $\sigma_{bs}$ emphasizes that improving the white noise reduces how long it takes the sensor to reach its ultimate precision but has little impact on the determination of its value.

Many published Allan deviation measurements only show the data up to averaging times of the bias stability at $\tau_{min}$.  This is unfortunate as the shape of the curve provides information on the scaling of the limiting noise.  Detailed study of systematic effects is challenging and rarely undertaken other than to eliminate them; however, appreciation of noise behavior informs design revisions to improve the measurement fidelity.

\subsection{Atomic Sensors \& the SQL}
The short term noise consists of the quadrature sum of the white noise terms from all sources $A=\sqrt{\Sigma A_i^2}$, which includes the quantum projection noise.  A common sensor optimization strategy focuses on reducing the largest of the non-quantum noise contributions until all $A_i$ terms are of similar magnitude--further reduction of an individual term has diminished impact in reducing $A$.  In cases where sources of technical noise have been sufficiently suppressed, an ideal atomic sensor reaches limitations due to fundamental quantum noise, the standard quantum limit.  
%At low frequencies, this is normally atomic projection noise, with a transition to photon shot noise at high frequencies.  
For most sensors, reaching the SQL would be a notable achievement, though rarely achieved even in the highest performance sensors.

Reaching the SQL could be particularly beneficial for electromagnetic sensors operated in quiet environments, where short-term sensitivity is valued over long-term accuracy.  For the highest precision laser-cooled microwave clocks, where atomic densities are purposely sacrificed to avoid collisional frequency shifts, the SQL is a very real obstacle.

Regardless of sensor performance, the SQL often represents an important benchmark:  whether the sensor is limited by technical noise or quantum noise informs the designer where improvements need to be made or, if there is a substantial SQL margin, may allow simplifications (i.e. adjustments to $N$, $T_2$, or gain) to be made without raising the fundamental noise floor above the technical noise.

%As the SQL depends on $N$ and $T_2$ (Eqn.~\ref{eqn:one}), efforts can be made to adjust its value.  In certain regimes, increasing the number of atoms will lower the limit, but this is only beneficial until an associated increase in collisions decreases $T_2$ or more importantly, raises the impact of systematic effects.  This balance, as outlined in Sec.~\ref{sec:sensitivity}, represents one the largest challenges of optimizing performance with regard to the SQL.

%There are some situations, for biomagnetic applications as an example, where the SQL is so far below laboratory noise levels that it has little relevance for individual sensor elements.  %The SQL may however recover its relevance for sensor arrays, see Sec.~\ref{sec:arrays}.

%Whether or not a particular sensor is operating at or near the SQL, the SQL serves as a useful benchmark.  If the SQL is far below the sensor noise level, that could indicate that the sensor could be made more compact, or lower power, or perhaps redesigned to a  lower gain configuration with less sensitivity to systematics.

Advanced quantum methods to surpass the SQL are discussed in Sec.~\ref{sec:QE}.

\subsection{Other Important Sensor Properties}
\paragraph{Bandwidth}
A typical sensor frequency response  is $S(f)=S(0)/\sqrt{1+(2\pi f T_2)^2}$, with the bandwidth being $b=(2\pi T_2)^{-1}$.  At frequencies above $b$, the measurement SNR degrades.  A common misconception is that feedback can enhance the bandwidth, but the gain enhancement at high frequencies is generally compensated by an increase in the effective noise level.  Note that since bandwidth and sensitivity are oppositely related to $T_2$, desirable improvements in sensitivity by reducing decoherence also degrade the bandwidth.  In this context, use of squeezing techniques can be valuable to increase the bandwidth without the usual degradation of the SNR at high frequency.

\paragraph{Dynamic Range}
Most sensors exhibit saturation when the sensed field becomes too large.  The dynamic range is typically the ratio of the largest field detectable without saturation to the smallest detectable field.  Again there is generally an inverse relationship between dynamic range and decoherence so that improving $T_2$ does not necessarily improve the dynamic range. Feedback is often helpful for increasing the dynamic range of a sensor.

\paragraph{Scale Factor Linearity and Stability}
For sensors like magnetometers and RF receivers, the primary goal is sensitivity, and the ability of the sensor to make an absolute field strength or calibrated measurement is less often exploited.  In contrast, the inertial sensors must not only have a high sensitivity, but  must be stable and have a linear response.  In principle, a model-based correction can compensate for a nonlinear scale factor, but could introduce long-term stability concerns.

Scale factor stability plays a particularly important role in sensor arrays in order to achieve a large common mode rejection ratio.  Inclusion of calibration fields and field gradients can mitigate scale factor (and offset) drifts, but clearly the system stability is a key property.

\paragraph{Size, Weight, \& Power (SWaP)} A dedicated sensor is often optimized for operation as a unit and may make desirable trades to accommodate a practical implementation.  The optimization of the system benefits from focusing on the ``long pole in the tent" where adjustments in the dominant contributors provide the largest gains in modifying the unit.  While further SWaP reductions may be possible, the ``cost" of doing so may play a role leading to SWaP-C as the overall considerations.

Ascribing to rigid SWaP metrics can be misleading.  For instance, creative evaluation criteria such as volume defined through water displacement provide less practical value than considering the envelope created by a regular parallelepiped.  In addition, steady state power draw for internal thermal regulation varies significantly depending on the external temperature.  Similarly, instantaneous power can exceed the steady state for short periods of time (ex: turn-on, pulsed operation etc.).  A single value may not accurately describe the needs for operation as a unit.

\paragraph{Orientation Dependence} 
Sensors of vector quantities come in a variety of measurement configurations.  Scalar sensors measure the magnitude of the vector, whereas ``vector" sensors measure one or more components of the vector.  Scalar measurements should remain invariant under changes in sensor orientation; however, depending on the configuration, they generally suffer from ``dead zones", directions in space where the sensor loses sensitivity.   Both scalar and vector measurements are subject to ``heading errors", a variation of the scale factor and/or offsets with orientation of the device.  The manifestation of dead zones and the sources of heading errors vary depending on the sensor type, but are always crucial to study and understand.

Several schemes exist to produce vector measurements from scalar sensors and vice versa.  These can be implemented with external field modulations, spatial arrays (Sec.~\ref{sec:arrays}), or other symmetry breaking techniques.  One potentially overlooked point worth emphasizing is that scalar sensors operating in a finite field are blind to small changes transverse to the principal direction of the field.  These sensors are most sensitive to variations that lengthen or contract the magnitude of the vector.

\paragraph{Power Cycling} Sensors operating outside the lab are often disconnected from a power source for some period of time.  A rapid start-up time can be highly desirable to achieve internal thermal equilibrium associated with operation. In addition, active vacuum pumps such as ion pumps cannot operate without power, so sufficient vacuum to perform a restart should be maintained through power interruptions.  Reduction of noticeable changes in sensor performance such as the ``retrace", fractional frequency shift in a clock from power cycling, are also desired.

\subsection{Sensor Arrays}\label{sec:arrays}
As quantum sensors become more compact and simplified, it becomes feasible to consider arrays of sensors.  Sensor arrays are particularly useful for spatial filtering, {\it i.e.} isolating a nearby source in the presence of a distant fluctuating background (or suppressing a strong nearby source so that fields from distant sources can be measured).  The larger the array, the more sources that can be isolated.  In the limit of a large number of elements, a sensor array becomes an imaging device that can produce a source map by inversion techniques.

Low noise and stable scale factor stability are essential for sensor arrays, both properties exploitable by atoms.  A key metric is the common mode rejection ratio (CMRR), which quantifies the ability of the array to suppress spatially uniform fields.  Scale factor stability is essential to maintain CMRR over time without degradation of the array performance.  A sensor array with scale factor stability problems must be actively calibrated with a known source, consuming valuable measurement resources.  

A well-matched sensor array with good CMRR may be able to overcome technical noise and operate at or closer to QPN limits. A recent gravity gradiometer demonstrated detection of a 147-kg lead mass through the vertical gradient of the vertical component $\Gamma_{zz}$ \cite{Janvier2022}.  Despite passive and active vibration isolation, the gravity vector was unresolved over the short integration time (1 h per position), while the common mode suppression in the gradient elevated the signal above the noise.  This feature has also allowed this and other gravity gradiometers \cite{Sorrentino2014} to perform at the QPN limit.  

Since the complexity cost of making sensor arrays scales approximately linearly with the number of elements, it is usually impractical to simply clone a successful complex single-channel sensing experiment into a sensor array.  Further simplification is often necessary.  An excellent example is specifically highlighted in Sec.~\ref{sec:magnetometer}.

A simple two-element array works best when the distance to the source is matched to the baseline between elements $d$.  For a field that scales as $\sim1/r^n$, a source at distance $D\gg d$ can be suppressed by $nd/D$.  A linear scaling in far-field suppression persists despite even high order magnetic dipole ($n=3$) or quadrupole ($n=4$) sources.  In addition, the differencing reduces the magnitude of the near-field source by a factor of $1-1/2^n$ for $D\approx d$, which can be significant for low order fields such as line currents ($n=1$).  More complex, large-element arrays require more sophisticated analysis, which can be addressed with signal processing tools like principal component analysis or machine learning.

Arrays can also be configured as vector gradiometers, specifically measuring  gradient tensor components $\partial_i F_j$.  For electromagnetic and gravity gradiometry where $\nabla\times F=0$ and $\nabla\cdot F=0$, a minimum of 8 well-chosen vector components at at least 3 locations can completely determine the uniform and gradient field components.  Accurate and stable calibration are essential for meaningful operation of gradiometers. 

%\cite{Lyu2022} is not at QPN, but does a nice analysis to optimize SNR.

\subsection{Breadboarding Value}\label{sec:breadboarding}%Breadboarding Strategy %High Value Breadboarding
Good sensor development practice nearly always requires an accompanying well-instrumented breadboard version, a key tool for rapidly or carefully exploring physics as well as implementation issues. The breadboard version can serve as a benchmark for the ultimate performance that can be achieved under ideal or nearly ideal conditions.  Sensitivity to external parameters can be evaluated and carefully documented for future designs. 
The breadboard system is particularly advantageous at evaluating systematic effects as the final quantum sensor has been stripped of helpful diagnostic tools.

A well-designed breadboard system incorporates elements to enable rapid turn-around, particularly on operations that would otherwise have long time constants. Ultra-high vacuum technology is one example where interior coatings \cite{Chu1998TheParticles} or a loadlock apparatus \cite{Kestler2024Optical-atomicH} can accelerate development.   

While it may be possible to leap from the breadboard demonstration to a finished design, sensor development often benefits from a series of intermediate prototypes with decreasing form factor and incremental integration into the final system.  These prototypes specifically freeze a selected subset of the degrees of freedom available on the breadboard as well as implement innovative simplifications in the approach towards the final sensor.  Testing the prototype against the benchmark breadboard system enables an appreciation for new insight into the physical system as some experimental imperfections have been amplified.

The number of these of iterations depends on the success of each step and the overall level of innovation pursued.  Throughout this progression, the benchtop implementation maintains its relevance.  Its flexibility enables supporting diagnostic experiments to identify shortcomings in the prototype design, or more importantly, tease out subtle physics or a systematic effect overlooked in the initial exploration.  Such novel findings should be expected and embraced rather than dismissed as experimental failure.  Even the most comprehensive breadboard experiments {\Edit may} inadvertently overlook a subtle design element that can be amplified in a prototype.  This symbiosis persists throughout the development process, particularly as advances in the prototype innovation produce initially unwelcome performance shortfalls, but serve as valuable lessons in sensor maturation.

%A particularly advantageous breadboard system incorporates elements to enable rapid turn-around such as the ultra-high vacuum in Ref.~\cite{Chu1998TheParticles}.  In addition, a loadlock apparatus can rapidly prototype chip scale devices for use in ultra-high vacuum \cite{Kestler2024Optical-atomicH}.

%{\Edit An intermediate step is the prototype, a more fully instrumented version of the sensor, which may retain many features of the lab system for studying systematics.}  %Some strategies in this process include exploitation of measurement symmetries or fundamental interactions to simplify or improve the measurement as well as clever reproduction of electro-optic operations in laser current modulations.}

%An initial prototype in a reduced form factor follows the benchtop system where its design specifically freezes a selected subset of the internal degrees of freedom.  Inevitably, the prototype enables an appreciation for new insight into the physical system as some experimental imperfections have been amplified. Iterations on this process focus on simplification requiring development of new tools, techniques, or configurations.

\subsection{Predictability \& Verification}\label{sec:sim-hardware}
A practical sensor experiences exposure to its environment and is often stripped of onboard diagnostics due to SWaP considerations.  This elevates the importance of simulating, testing, and validating the packaged sensor, under a range of external conditions.

Like the modeling of the atomic physics (Sec.~\ref{sec:evolution}), sophisticated commercial software simulation packages such as COMSOL Multiphysics, Anysys, Algoryx, Zemax etc. aid in understanding the behavior of the sensor package.  Such tools can evaluate the impact of dielectric and magnetic permeability of the housing, optical deflections under dynamics or vibration, and thermal coefficients.  These material properties can be connected to the physics model established previously to determine the contribution to the sensor output.  Extensive hardware testing confirms that the sensor behaves as predicted.  Tests include reversal of the sensor in a known field to evaluate bias, application of known fields and modulations to determine scale factor, and Allan deviations to evaluate long-term stability.  In addition, testing performed over environmental ranges provides valuable environmental coefficients that can be applied as small corrections at the system level as needed.  All performance limits should be understandable from the modeling of the instrumentation and physics. 

\section{Atomic Sensor Development}\label{sec:development}
The process of transitioning from expensive, complex apparatus with extensive diagnostics run by experts (i.e. quantum sensing) into a self-contained, cost-effective device for non-expert use ({\Edit i.e. }quantum sensors) presents challenges beyond physics (Sec.~\ref{sec:atomSensors}) and implementation (Sec.~\ref{sec:implementation}). Much of Section~\ref{sec:implementation} addressed important issues such as SWaP, complexity cost, sensor metrics, predictability and verification, and breadboarding that arise in transitioning quantum sensing experiments into practical quantum sensors.  Here{\Edit ,} we note some large-scale issues associated with the development of atomic quantum sensors.%  Successful sensor implementation involves contributions from a variety of competing organizations, over decade timescales.  It is very difficult for a single organization to develop a quantum sensor technology in isolation.  Similarly, attempts to fast-track sensor development by simply scaling down a successful quantum sensing experiment rarely produce a quality sensor.

\subsection{Community Research Efforts}\label{sec:community}
Efforts to build atomic sensors usually involve a mixture of universities, government labs, and commercial developments, whether collaboratively or in competition.  It is rare for a single entity to independently create a series of innovations leading to a successful sensor.  More often, a community-driven effort plays a substantial role in accelerating innovation.

An excellent example of multiple groups advancing sensor development is the  SERF magnetometer. 
%which has recently been established in array of miniature sensors \cite{Alem2023}.
The original SERF demonstration in Ref.~\cite{Allred2002} was implemented on an optical table using a hand-blown vapor cell heated in an oven using forced hot air contained within multiple layers of magnetic shields.   Two independent lasers {\Edit implemented}\strike{provided}  optical pumping and probing functions with a water-cooled Faraday modulator providing polarization modulation on the probe beam \strike{producing}{\Edit to produce} measurements along a single vector projection of the magnetic field. Over time, many aspects of this implementation evolved.  For example, lower technical noise heating methods such as electrically resistive \cite{Kornack2007} or optically heating with color glass \cite{Mhaskar2012} were introduced.  A photo-elastic modulator (PEM) provided a lower drift modulation over the Faraday modulator \cite{Smullin2008}. Field modulations eventually replaced the need for light modulations \cite{Li2006parametric}.  These and many other innovations are implemented by and further advanced in a variety of commercial magnetometer efforts \footnote{{h}ttp://www.twinleaf.com; http://www.quspin.com}.

SERF sensor arrays were initially explored in a common vapor cell with a detector array sampling multiple detection zones on a mm-scale baseline. \cite{Kominis2003,Xia2006}.  Following this, compact sensors with cm-scale baselines were constructed and implemented into simple 4-channel arrays \cite{Wyllie2012OpticalMagnetocardiography,Wyllie2012MagnetocardiographyArray}.  In parallel, efforts to miniaturize other magnetometers with microfabrication techniques \cite{Schwindt2004Chip-scaleMagnetometer,Schwindt2007Chip-scaleTechnique,Shah2007SubpicoteslaCell} led to miniature sensors operating at zero field \cite{Ledbetter2008PNAS}.  These efforts benefited from prototype demonstrations that demonstrated simplifications to the sensor architecture such as pumping and probing functions with a single laser beam \cite{Shah2009Spin-exchangeLight} {\Edit or independent, co-propagating pump and probe beams \cite{Johnson2010}}. Commercialization followed the developments in basic research laboratories, see Sec.~\ref{sec:magnetometer}. 

The two-decade SERF development process began with a variety of academic and governmental research efforts openly disseminating and sharing their fundamental and practical advances.  This is in contrast to common practice in industrial laboratories which generally do not share their innovations, even when the development costs are directly publicly funded. It is hard to see how secrecy benefits the necessarily long-term development of atomic sensor technology.

\subsection{Development Retrospective}\label{sec:retrospective}

Quantum sensor development can succumb to a temptation to directly co-opt a design from a fundamental science measurement in an academic laboratory and reproduce the science experiment in a box.  Here, engineering management processes generate specification requirements closely resembling the laboratory hardware systems and subsystems and modeling replaces in-house breadboarding of the system.  Following these specs, custom hardware is designed, assembled, and integrated into a prototype unit.  Commonly, the resulting hardware falls short in areas such as dynamic range, orientation independence, or thermal stability etc. as they were irrelevant for the original laboratory system.

The rigid process described above is a direct consequence of a high resource investment over a short development time-frame that desires revolutionary progress under a single funded effort.  The devil is in the details to achieve the performance capabilities of atom-based sensors, details that are quickly abandoned under the pressure of aggressive development time-frames.  While a bold challenge can motivate innovation,  the artificially shortened execution timeline often stifles creativity and de-emphasizes further discoveries. %The appreciation of which led to the initial success in the scientific advancements that preceded it.  

A more natural and technical advancement driven process allocates the same fiscal resources over longer horizons, enabling a series of incremental advancements across research groups to provide a foundation for revolutionary insights.  This measured approach introduces compatibility with quantum workforce development, educating traditional engineers to support a quantum technology by sustaining workforce attention on quantum problems of relevance.

%Before continuing, we wish to acknowledge the important milestones provided by ``lab in a box".  Such intermediates that exit the laboratory show promise for a next generation and confirm that additional fundamental physics innovations are clearly warranted.  It is rare for any technology of sufficient complexity to advance from a ``proof-of-concept" directly to the final device without a series of intermediate steps and supporting innovations.  

%{\Edit A driving question throughout the development process asks how to create physical conditions that support measurements with the atom advantage (Sec.~\ref{sec:atom_adv}).}

\section{Environmental Challenges}\label{sec:environChallenges}
Quantum sensing in research laboratories often relies on highly controlled environments.  In contrast, practical quantum sensors must  operate in the presence of uncontrolled electromagnetic fields (and their gradients), platform motion, environmental temperature, humidity etc. In the following{\Edit ,} \strike{we give} some basic information on these issues and some examples of how they are dealt with for sensing in harsh environments {\Edit is given}.

\subsection{Environments}\label{sec:environments}
\paragraph{Electromagnetic Backgrounds}  It is common to compensate for earth's $20-80~\mu$T magnetic field by adding shimming fields or placing experiments inside passive magnetic shields.  A fielded sensor, in contrast, must operate through dynamic and unpredictable environmental noise.  The sensor platform may be moving, causing the earth's field to change orientation; communications equipment, motors, or ferrous objects may be nearby; noisy power lines and poorly grounded circuitry inject high levels of noise into the environment.  Fields from power lines are particularly difficult to avoid due to the $\sim 1/r$ scaling of magnetic fields from line currents (10-20~nT at distances of 30~m).  Even some of the most remote locations have substantive electromagnetic backgrounds--ocean surface waves produce fields of magnitude 100~pT-1~nT depending on surface conditions \cite{Avera2009}.  Earth's B-field itself presents $\sim 100~$fT/$\sqrt{{\rm Hz}}$ with a significant $1/f$ contributions below 1 Hz and is subject to a variety of time-varying geophysical phenomena from lightning strikes to solar weather \cite{Constable2016}.

%\Thad{Is the intent to include typical responses to these challenges?}

Clocks and inertial sensors generally address these problems  with magnetic shielding and internal coils to set the bias field.  This may also work for electromagnetic sensing of small nearby objects that can be included inside a shield. Electromagnetic sensing of fields from distant objects is necessarily unshielded and thus faces severe challenges.  Feedback may be used with local shimming coils to enhance linearity and dynamic range.  Gradiometry using sensor arrays (Sec.~\ref{sec:arrays}) has the potential to isolate faint sources in the face of large background fields.

\begin{table}
    \centering
    \begin{tabular}{c|ccc}
        Temperature Grade & \multicolumn{3}{c}{Range}\\
        \hline
        Commercial & $0^\circ$C & to & +85$^\circ$C\\
        Industrial & $-40^\circ$C & to & +100$^\circ$C\\
        Military & $-55^\circ$C & to & +125$^\circ$C\\
        Low Earth Orbit (LEO) & $-65^\circ$C & to & +125$^\circ$C\\
        Geosynchronous Orbit (GEO) & $-196^\circ$C & to & +128$^\circ$C\\
    \end{tabular}
    \caption{Typical environmental temperature ranges.}
    \label{tab:tempranges}
\end{table}

\paragraph{Temperature Variations} 
Fielded sensors encounter much greater temperature variability than in the typical research laboratory.
Some common examples are included in Table~\ref{tab:tempranges} where satellites experience the largest temperature swings.  More stable environments exist in downhole drilling at $>200^\circ$C or subsea applications where surface temperatures vary with latitude but approach a uniform $\sim 4^\circ$C temperature at depth.  To adapt to these challenges,  the sensor temperature is often stabilized by a thermal feedback loop to a set point slightly above the highest expected external temperature. 

Most atomic sensors include internal thermal control that allows the sensor to operate at a temperature that is convenient for the sensor and its components, being particularly cognizant of vapor cell or vacuum enclosure, the electronics, and the lasers.  In particular, competing temperature operating points for lasers and heated cells are often in conflict, which is driving development of commercial high temperature laser sources.

\paragraph{Platform Motion} Fielded atomic sensors experience  accelerations, rotations, and vibrations rarely present in the laboratory.  These can introduce deflections in the mechanical structure (optical paths, field coils etc.) or laser frequency shifts (Doppler etc.) unless properly engineered.  In addition, the dynamic range of a cold atom based sensor can be restricted due to atoms colliding with the wall following trap release.  %For space-based systems in freefall, cold atoms can be temperature-limited with $>10$~s coherence times; however, evaporated atoms do not fall away from the trap as in a terrestrial experiment and produce background signals in detection \cite{Wililams - Cal reference}.
In some cases such as navigation, accurately measuring platform motion is the goal.  An inertial navigation system (INS) determines change in position without external references by accurately measuring platform motion over all six degrees of freedom from known initial conditions (position and orientation).  The underlying integrals determining change in position compound any sensor errors (statistical and systematic) into a growing position uncertainty that can predicted through Kalman filters \cite{Gelb1974AppliedAnalysis}.  Table~\ref{tab:inertial_metrics} provides relevant sensor performance metrics for mature inertial navigation instruments operating over temperature and dynamic range with $\sim 1$~kHz bandwidths. 

\begin{table}
    \centering
    \begin{tabular}{cc|ccc}
\multicolumn{2}{c|}{Inertial Navigation}    & Tactical  & Navigation    & Strategic\\
\multicolumn{2}{c|}{Parameters}                & Grade     & Grade         &  Grade\\\hline
\multirow{3}{*}{\rotatebox[origin=c]{90}{Accel.}}
                                 &Noise $(\mu g/\sqrt{{\rm Hz}})$ & 50-75     & 3-10          & 0.1-0.5\\
&Bias $(\mu g)$                  & 500-1000  & 10-100        & 0.1-10\\
&Scale Factor (ppm)              & 1000-3000 & 10-50         & $<1$ \\\hline
\multirow{3}{*}{\rotatebox[origin=c]{90}{Gyro.}}
                                 &Noise $(^\circ/\sqrt{{\rm hr}})$ & 0.1-0.5  & 0.0005-0.002  &$<0.0001$\\
&Bias $(^\circ/{\rm hr})$        & 5-50      & 0.003-0.02    & $<0.003$\\
&Scale Factor (ppm)              & 500-1500  & $<15$           & $<1$ \\
%\hline
%\multicolumn{2}{c|}{Typical Cost/Unit}               & \$200-\$2k& \$20k         & \$250k-\$1M 
    \end{tabular}
    \caption{Relevant metrics for accelerometers and gyroscopes by grade for inertial navigation.  Approximate ranges reflect dependency on precise path traveled to determine position uncertainty. A typical navigation-grade system common on manned and unmanned aircraft have a typical center error probable growth rate of  1 nautical mile per hour.  Bias considers the response of the sensor over environmental conditions.}
    \label{tab:inertial_metrics}
\end{table}

Inertial navigation represents a challenging, long-term goal for atom-based sensors that capture the advantages of atoms if dynamic range and volume considerations can be addressed.  This is likely to occur in navigation-grade and strategic-grade systems as a large number of mass-producible, chip-scale technologies already dominate tactical grade market.  Map-matching to geophysical features such as gravity \cite{Kohler2023} and magnetic anomalies \cite{Canciani2017,Canciani2022} present near-term navigation impacts, again if the impact of the dynamic platform can be managed.

 %Navigation-grade and strategic grade metrics serve as targets for evolving atom-based sensors that capture the advantages of atoms if dynamic range and volume considerations can be addressed.  A large number of mass-producible, chip-scale technologies already dominate tactical grade market.  While inertial navigation remains a challenging, long-term goal, single-axis gravimeters and gradiometers are likely to advance the technology as well as serve as navigation aids such as map matching to geophysical features, gravity \cite{Kohler2023} and magnetic \cite{Canciani2017,Canciani2022}.

\paragraph{Other Considerations} Radiation, humidity, pressure, etc. are also relevant.  These can often be mitigated by hermetic sealing of the sensor enclosure and selecting radiation hard materials. {\Edit Space qualification is particular challenging, often introducing re-evaluation and extensive testing of all components.}

\subsection{Environmental   Adaptation Examples}\label{sec:env_advances}
There is a  strong synergy between compact atomic sensors and fundamental science in challenging environments. For both types of applications, careful consideration of complexity vs. performance are required, plus a large premium on remote operation and (sometimes) unskilled operation.  Precision instrumentation benefits from robustness and simplicity no matter the environment.  In this respect, fundamental research in harsh environments is very likely to result in advances in atomic quantum sensors, and vice versa.  Fundamental science experiments in harsh environments still contain internal diagnostics to address systematic effects but serve as an accelerator for supporting technology development.

%Naively one divorces the goals of fundamental science experiments and practical applications in challenging environments, but precision instrumentation benefits from robustness and simplicity no matter the environment.  Operation of quantum sensors in remote environments such as space and subsurface places a premium on unskilled operation with minimal intervention.  Atomic clocks have operated in space to enable the GPS constellation for terrestrial navigation.  

For example, complex cold atom systems including atom interferometers have been engineered to survive drop towers \cite{Muntinga2013InterferometryMicrogravity} and sounding rockets \cite{Lachmann2021} at a variety of terrestrial microgravity facilities \cite{Raudonis2023MicrogravityExperiments}.  These developed experiments operate reliably with minimal intervention and demonstrate that complex cold-atom hardware can be matured for use outside the laboratory.  An outgrowth of these experiments is the Cold Atom Laboratory (CAL) on the international space station (ISS) \cite{Aveline2020}.  This was installed by non-specialists and has been operated completely remotely by design \cite{Frye2021}.

Optical lattice clocks have achieved fractional frequency uncertainty at $10^{-18}$ or below{\Edit.  Such clocks are }capable of accurately measuring the gravitational potential at cm scales{\Edit, where more recent work has achieved measurements on mm-scales \cite{Zheng2022,Bothwell2022ResolvingSample}}.  A local measurement of the gravitational redshift $\Delta z g/c^2$ is particularly valuable for monitoring crustal deformations associated with earthquakes or volcanic activities where height changes in the geopotential $\Delta z$ are more significant than mass redistribution impacting $g$ \cite{Takamoto2022}.  Such monitoring requires transport of some the highest performance clocks.  A Sr optical lattice clock has been housed in an air conditioned trailer \cite{Koller2017} and a pair of Yb clocks has been built and compared locally \cite{McGrew2018}.  Continued maturation of portable systems has led to side by side and remote comparisons of clocks at different geopotentials \cite{Grotti2018, Grott2024}.

The freefall and stationary environments discussed above offer harsh but relatively predictable conditions.  Sensor operation on  a surface ship at sea requires operation in a highly variable environment.  Recently, several nearly identical optical clocks based on an iodine vapor and fiber frequency comb operated simultaneously over a 20~day evaluation period \cite{Roslund2024OpticalSea}.  By design, these clocks are first order insensitive to platform motion and leverage mature telecom laser components at 1064~nm and 1550~nm.  Hydrogen maser-like performance of $5-6\times10^{-14}/\sqrt{\tau}$ was maintained in {\Edit two }\strike{ the }35~L \strike{PICKLES and EPIC }clocks while a similar further simplified \strike{VIPER }clock maintained $5\times10^{-13}/\sqrt{\tau}$.  Testing exposed the clocks to accelerations, \strike{of $\sim1$~m/s$^2$, rms }vibration{\Edit s}\strike{ of 0.03~m/s$^2$}, {\Edit and }rotations\strike{ of 1-6$^\circ$ at 1-3$^\circ$/s}.  An air-conditioned \strike{Conex }box limited some outside influences, but still introduced {\Edit modest }\strike{2-3$^\circ$C }thermal and \strike{4-5\% }relative humidity variations.  Overall, the clocks performed similarly to laboratory tests, immune to ship motion but still subject to small changes correlated with the day-night temperature/humidity cycles.

An example of a robust commercial sensor is the  Chip-Scale Atomic Clock (CSAC).  The CSAC is a very different system in that it has traded sensitivity for SWaP to the extreme.  The Microchip SA.45s CSAC provides $3\times10^{-10}/\sqrt{\tau}$ over a temperature range of $-40^\circ$C to $+80^\circ$C.  This is achieved in an 8~mm$^3$ microfabricated  cesium vapor cell containing buffer gases, heated to 90-95$^\circ$C, and interrogated with a single, low-power VCSEL.  The entire $\sim$15~cm$^3$ sensor consumes 125~mW with only 10~mW of the power budget for the elements in Fig.~\ref{fig:simple_sensor} \cite{Lutwak2007}.  Based on the internal thermal setpoint, it is difficult to address the temperature ranges in Table~\ref{tab:tempranges}:  increased temperature detrimentally broadens the line due to collisions and reduces signals {\Edit as} the vapor becomes opaque.  A recent innovation replaces Cs with a K-Cs mixture to translate the optimized vapor conditions to a 115$^\circ$C setpoint, enabling operation above 105$^\circ$C \cite{Ha2024}.  The mixture synthetically reduces the cesium vapor density by the mole fraction of the mixture ($X_{Cs}=0.2$) following Raoult's Law.  By mixing with a secondary species with lower vapor density, the Cs-Cs collisions remain the dominant line-broadening factor, \strike{albiet} {\Edit albeit} at an increased temperature.

\section{Emerging Supporting Technology}\label{sec:support_tech}

Atomic sensors rely heavily on laser, electro-optical, specialized packaging, and electronic technologies that enable high-performance, robust, and compact atomic sensors.   Here we describe three examples, among many, of emerging supporting technologies that are key for future developments:  photon manipulation tools\strike{ (PICs and advanced VCSELs)}, cold atom vacuum systems, and miniature vapor cells.  Further maturation of these tools will accelerate atomic sensor development.

%The type of platform has significant implications on the final volume of the sensor, based largely on the complexity of contents of the system (Sec.~\ref{sec:platform}).  A single sensor based on a vapor can be managed in $<500$~cm$^3$ (or significantly less depending on the measurement) whereas a cold atom systems typically struggles to achieve a $\sim 100$~L volume.

\subsection{Photon Manipulation Tools}\label{sec:integrated_photonics} A recurring challenge for quantum sensors is effectively implementing optical infrastructure for state preparation, evolution, and detection functions into a self-contained sensor unit.  Tabletop free-space optical systems for cold atoms have been transformed into ruggedized, smaller-form factor integrated racks \cite{Schmidt2011AExperiments} or miniature optical benches \cite{Pahl2019CompactMicrogravity}, even enabling such systems to be dropped from towers \cite{Schkolnik2016ARocket}.  A recent complementary approach leverages mature, optical fiber-components developed for the telecom industry \cite{Diboune2017Multi-lineInterferometry,Jiang2022LowInterferometry}, though these systems exhibit low wall-plug efficiency due to losses associated with doubling required to reach the usual alkali-metal wavelengths.  Overall, the volume of the optical generation, manipulation, and delivery systems can easily dominate the sensor form factor.

\subsubsection{Photonic Integrated Circuits (PICs)}
A photonic integrated circuit (PIC) incorporates more than one electro-optic function into a single, compact platform or substrate.  In principle, any number of lasers, modulators, detectors, switches, isolators etc. interconnected by high-density waveguides can be integrated into a single cm-scale chip \cite{QEDC-PICS23}.  In effect, lithographically-defined waveguides replace free-space propagation or fiber coupling/routing in a dramatically reduced footprint.  For atom-based sensors, this SWaP advantage is accompanied by an increase in robustness: integration builds stability by eliminating opportunities for optical misalignment between components.

%through integration as each additional optical element (free space or fiber) represents another opportunity for mechanical misalignment. 

%this represents a dramatic reduction in SWaP, but more importantly, such optical integration leads to robustness where each additional optical element (free space or fiber) represents another opportunity for mechanical misalignment. 

A frequently promoted vision offers replacement of the entire sensor optical system (free-space or fiber) with an equivalent chip.  This is challenging for PICs as no single material platform can simultaneously incorporate electrically-injected photon generation (laser), electro-optic manipulation (modulators etc.), and low loss propagation (waveguides) as a monolithic structure.  Complex processing of hybrid (disparate materials coupled through high-precision placement) or heterogeneous (mixing disparate materials at the wafer level) structures is required.  
Promising materials for native atomic wavelengths include GaAs \cite{Dietrich2016GaAs} for photon generation and silicon nitride Si$_3$N$_4$ \cite{Blumenthal2018SiliconPhotonics} along with thin-film lithium niobate \cite{Zhu2021IntegratedNiobate} for modulators.  A realizable, near-term approach focuses on creating an integrated chip or chips that address(es) several key optical functions in conjunction with a small number of free-space optics.

%In addition, materials that natively support the relevant atomic wavelengths are still developing, where silicon nitride Si$_3$N$_4$ \cite{Blumenthal2018SiliconPhotonics} and thin-film lithium niobate (TFLN) \cite{Zhu2021IntegratedNiobate} show great promise in supporting the necessary 

%These various functions can be incorporated in a monolithic (defined through the same material), hybrid (disparate materials coupled through high-precision placement), or heterogeneous (mix) structure.  integrated in

%More importantly, such optical integration leads to robustness--each additional optical element (free space or fiber) represents another opportunity for mechanical misalignment. 

%{\Justin Where does QED-C Reference go?} \cite{QEDC-PICS23}

Quantum sensors represent a small niche among other more prominent PIC applications such as lidar, optical computing \cite{McMahon2023TheComputing}, and communications.  As such, PIC development for quantum sensors is not market driven, but researcher driven, mainly at universities or corporations under government sponsorship.  Here, researchers have produced proof-of-concept component technologies such as non-magnetic optical isolators \cite{Tian2021Magnetic-freeIsolator,Sohn2021ElectricallySplitting,White2023IntegratedIsolators}, offset phase-locked lasers up to 16~GHz at 1550 nm \cite{Arafin2017HeterodyneModulators}, and narrow clock lasers with fractional frequency noise $<10^{-13}/\sqrt{{\rm Hz}}$ \cite{Loh2015Dual-microcavityLaser} or sub-Hertz fundamental linewidth \cite{Gundavarapu2019Sub-hertzLaser} that have clear opportunities for use in sensors if the relevant wavelength, optical power, and narrow linewidth can be achieved.  %Other corporate achievements and capabilities may remain unpublished.

Photon manipulation structures in PICs (ring resonators, circulators, interferometers etc.) differ from the traditional electro-optical and opto-mechanical devices (AOM, EOM, shutters etc.) commonly exploited in AMO science.  As such, a different set of nuanced behaviors (residual amplitude modulation, extinction, sidebands etc.) in these tools can introduce a systematic effect or instability in a sensor application.  There is therefore enormous value in rigorous testing of PICs in sensing contexts \cite{Hummon2018PhotonicInstability,Chauhan2024TrappedLaser}.  In addition, in designing a PIC to support a sensor, it may be insufficient to rely on communicating a ``requirements" list to a photonics engineer at a university or company with a foundry.  Instead, interdisciplinary collaboration between sensor and PIC specialists should define appropriate on-chip structures matched to a sensor configuration that fully leverages the emerging PIC toolkit.  Such collaborative efforts will require sustained investment to accommodate the PIC fabrication cycles (3-6~month typ.).

Once a suitable PIC has been developed, an open challenge remains its integration into the larger system, particularly in the coupling to fibers, other PICs, or free space.  Edge coupling from the sides of the chip as well as grating coupling normal to the chip must be coordinated with the sensor design; further maturation of the input/output coupling is of benefit as $\gg 3$~dB losses per interface is not uncommon.  In addition, the lower SWaP introduces a chip-level thermal sensitivity (internal or external), requiring proper heat-sinking of a small thermal mass.  
%Challenges: lower mass leads to thermal sensitivity, integration into larger systems, external coupling to fibers, other PICs, or free space. - ultrahigh vacuum \cite{McBride2024}

It is clear that even with the maturation of advanced PIC technology, conventional optical components will still retain a prominent role in practical sensors.  To address SWaP, development of creative optical designs using a minimum number of existing tools combined with design one-upmanship provide great benefit.  For example, it is possible to implement all atom interferometer operations with a single laser and small numbers of electo-optics.  Wu {\it et al.} \cite{Wu2017MultiaxisTrap} demonstrate operation using a single diode laser, fiber EOM and 5 AOMS while L{\'o}pez-V{\'a}zquez {\it et al.} \cite{Lopez-Vazquez2023CompactGravimeter} reduce the electro-optics to one EOM and one AOM.  Similarly, Theron {\it et al.} \cite{Theron2014NarrowInterferometry} achieve operation using a single, telecom fiber laser and two phase modulators doubled to 780 nm.  %{\Edit This kind of component reduction combined with design one-upmanship fosters continued innovation to support the quantum sensor ecosystem.}

%PIC System example  - on chip SBS laser stabilize to a 3~m cavity, $^{88}$SR$^+$ ion clock and qubit operations \cite{Chauhan2024arXiv}, \cite{McBride2024} \cite{Hummon2018PhotonicInstability}

\subsubsection{Advanced VCSELs}
Useful sensors can be enabled with a laser and a small number of optics, without requiring the capabilities of a PIC.  Examples include the CSAC and the magnetometer array (Sec.~\ref{sec:magnetometer}) which leverage a single Vertical Cavity Surface Emitting Laser (VCSEL) as the light source.  The VCSEL provides key advantages over edge emitting diodes such as high efficiency photon production with sub-mA threshold current, circular spatial mode profile, fast modulation ($\gg1$~GHz), and integrated optical isolation due to a high-reflectivity output mirror.  Unfortunately, commercial VCSELs for relevant atomic transition wavelengths produce only $\sim1$~mW of optical power with 50-70~MHz linewidths \cite{Zhou2022VCSEL}, traditionally limiting their use to small form factor vapor cell sensors (clocks, magnetometers, etc.).  If advanced VCSELs with 15~mW of optical power and $\sim1$~MHz linewidths could be produced in conjunction with their other attractive properties, a greater number of applications could be addressed.  For instance, a MOT with $10^7$ atoms could be produced with high optical efficiency from a single current-modulated chip, a collimating optic, and a small number of reflective optics much more elegantly than some integrated photonic approaches \cite{Ropp2023IntegratingChip}.  A recent, research-grade 3~mW VCSEL with $<2$~MHz linewidth shows promise \cite{Huang2022VCSEL}.

%metasurface as beamexpander from PIC waveguide mode \cite{McGehee2021}\\
%Integrated photonic MOT \cite{Isichenko2023}\\
%Charge noise limitation of flicker type noise \cite{Zhang2023}\\

\subsection{Cold Atom Vacuum}\label{sec:vacuum_innovation}
Laser cooling methods isolate atoms from interactions with the walls through momentum state preparation to literally suspend them in space.  This can only be achieved under specific UHV conditions that have been next to impossible to demonstrate in the kind of simple glass cells commonly used in warm vapor experiments. In the case of neutral atom trapping in a MOT, background densities of the trapped species of $\sim10^8$/cm$^3$ (room temperature of \strike{K} {\Edit potassium}) and background pressures of $10^{-8}-10^{-9}$~Torr provide optimal performance, but atoms can {\Edit only} be trapped up to background pressures of $10^{-7}$~Torr. Ion traps typically operate at $10-100\times$ lower background pressures to enable longer trap lifetimes and require substantially fewer trapped species for loading. These conditions have traditionally been implemented in laboratory systems using a stainless steel vacuum chamber, glass windows, source of atoms, ion pump, and passive titanium sublimation pump or other chemical getter(s) \cite{Lewandowski2003}.  Following an UHV bake-out, the established approach provides a reliable system that balances continuous background contamination from both materials out-gassing and atmospheric helium permeation through glass.

%More recent vacuum - Farkas2010APL, Rudolph2015_NJP

%Advances in vacuum techniques have achieved atom-trapping conditions in small form factors with some configurations becoming commercially available. Many of these designs follow a direct miniaturization approach, incorporating similar features to the traditional laboratory systems such as separate atom loading and measurement regions (Fig.~\ref{fig:vacuum_innovation}a).  One particular challenge in direct miniaturization to support sensors is the decreasing distance between the ion pump magnet and the measurement region.

%As shown in Fig.~\ref{fig:vacuum_innovation}a, the basic cold-atom trapping vacuum approach has persisted since the earliest laser cooling experiments.  {\Justin variation on theme} Here, separate regions isolate each experimental step from initial cooling in a MOT followed by transfer to a dedicated science chamber.  The highest vacuum is required in the science chamber and achieved by an active ion pump and getters.  Many designs incorporate these basic features, albeit in smaller form factors. Direct miniaturization leads to challenges in practical quantum sensors, particularly with decreasing the distance between the ion pump magnet and the measurement region.  
 
\begin{figure}
    \centering
    \includegraphics[width=1\linewidth]{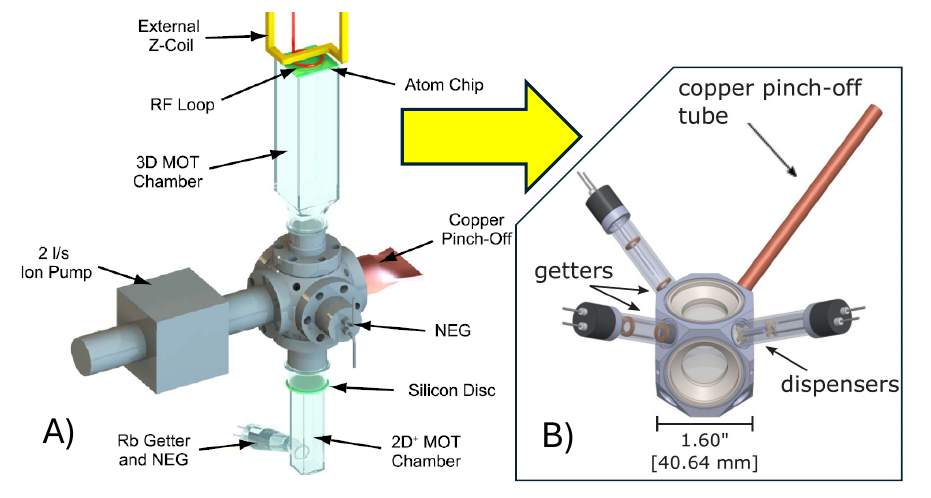}
    \caption{A) A small-scale UHV vacuum system containing active pumps and separate trapping and measurement regions reproduced from Ref.~\cite{Farkas2010} B) Passively-maintained, compact UHV system reproduced from Ref.~\cite{Little2021}.}
    \label{fig:vacuum_innovation}
\end{figure}

%A new imagining of the approach modulated the vapor pressure in a single vapor cell between the established ideal loading and clean measurement conditions on a 100~ms timescale \cite{Dugrain2014}.  A further innovation greatly simplifies the measurement cycle by replacing pressure modulation with recapturing atoms in a MOT at the same location to enable faster measurements at $\sim100$~Hz \cite{McGuinness2012}.  

Miniature components have been developed to address technical challenges in small-scale designs including sourcing atoms, leak mitigation (UHV bonding and coefficient thermal expansion matching), gas permeation (primarily He), and outgassing (H$_2$, CO, noble gasses etc.), see Appendix.  Small scale designs have been achieved incorporating similar features to the traditional laboratory systems such as separate atom loading and measurement regions (Fig.~\ref{fig:vacuum_innovation}a).  Much of this advancement has focused on supporting a MOT, but not fully addressed the more complex needs of a sensor, particularly mitigating decoherence in the measurement.  The decreasing distance between the ion pump magnet and the measurement region in a miniature system is a challenge.

To support high-performance measurements, the vacuum solution must simultaneously provide (1) alkali vapor regulation at $\sim10^8$~cm$^3$ ($\sim10^{-8}$~Torr), (2) background gas maintenance $\lesssim 10^{-8}-10^{-9}$~Torr (including noble gases), (3) limited decoherence (magnetic field interference etc.), and (4) unrestricted optical access for the sensor optical configuration.  A passive solution to these conditions that can be produced at the cm-scale has the potential to become the universal standard, particularly if it can operate at an internally-stabilized elevated temperature (Fig.~\ref{fig:vacuum_innovation}b).

Such a solution is likely forthcoming based on recent advancements.  Chambers made from carefully selected materials with low-helium permeability offer the potential to operate without an ion pump once purged and sealed.  Notable materials include an aluminosilicate glass compatible with anodic bonding \cite{Dellis2016}, a sintered alumina chamber with glass-ceramic windows \cite{Burrow2021}, and a titanium chamber brazed to sapphire windows \cite{Little2021} (originally supported ion traps \cite{Jau2012Low-powerPackage,Schwindt2016AClock,Scherer2018AnalysisClock}).  These approaches offer a simpler and more compact structure than a similar anodically bonded glass and silicon chamber which necessitates an appendage to support a spatially-separated ion pump \cite{McGilligan2020LaserPlatform}. As an added advantage, these materials reduce eddy currents \strike{assicauted from} {\Edit associated with} rapid magnetic field switching in neutral atom trapping applications where even the modest conductivity and magnetic permeability of titanium offers substantial reduction over stainless steel.  Refs.~\cite{Burrow2021,Little2021} demonstrate operation without an ion pump where in the case of Ref.~\cite{Burrow2021}, the ion pump is initially included, but pinched off.  Long-term operation on the scale of years has been demonstrated.  {\Edit A similar approach for ion trap packages have maintained state of the art clock stability after only passive pumping for 10 years \cite{Thrasher2025}.}

Active atom sources to produce alkali metal at the desired background density for trap loading are common to address external temperature swings or elevated temperature operation (Sec.~\ref{sec:environments}). Numerous approaches for controlled alkali sourcing have been developed for reduced SWaP \cite{Fortagh1998,Gong2006ElectrolyticCells,Scherer2012CharacterizationSensors,Bernstein2016SOLIDSENSING,Kang2017ASource,Kang2019Magneto-opticSource,Rusimova2019AtomicApplications,Kohn2020CleanGraphite,McGilligan2020Source}.  A passive atom source would be preferred as these active approaches can contaminate the vacuum and compete for optical access on the chamber surface.  An equilibrium vapor matched to the sensor operating temperature should be possible \cite{Ishikawa2017,Ha2024}.  A final challenge would be integrating these aspects with a UHV bake-out and suitable seal such as membrane-closure or cold-weld, or alternatively performing operations {\Edit via} in-vacuum encapsulation \cite{McGilligan2022}.

\subsection{Miniature Vapor Cells}\label{sec:vapor_cells}
A typical hot atom sensor includes a sealed glass chamber with a small droplet of alkali metal inside.  The cell is in thermal contact with an oven that may be heated by a variety of methods (electrical, optical, convection etc), and whose temperature sets the equilibrium alkali-metal density.  Since the vapor pressure inside the cell must be spatially uniform, the inevitable temperature gradients across the cell cause the alkali-metal to form a droplet at the lowest temperature point, with a thin film forming elsewhere such that the flux of atoms emitted from each small surface area of the film is equal to the flux of atoms incident on that same area.  It is important to keep in mind that even with optical or anti-relaxation coatings on the cell walls, a thin film of alkali-metal forms in order to maintain an equilibrium between the condensed phase and vapor phase.  Preparation of a cell requires some sort of interior surface cleaning, metal deposition that may or may not include surface coating or buffer gas, and final seal.

For sensors greater than 1 cm in diameter, it is usually easiest to form blown thin-walled spheres or attach optical windows to cylinders through various methods.  The desire for miniaturization has spurred a variety of cell construction techniques such as anodic bonding of glass to etched silicon wafers \cite{Shah2007SubpicoteslaCell}.  Millimeter-scale cells  as well as the extreme case of sub-micron waveguides \cite{Skljarow2020} have been achieved.  Recent specialized glass cells include micro-glass-blown atomic vapor cells \cite{Noor2020DesignCells} and anodic-bonded aluminosilicate glass that limits atmospheric helium permeation \cite{Dellis2016}.  Recent fabrication techniques include cutting cells with hot wires \cite{Laliotis2022SimpleSpectroscopy}, femtosecond machining inside a fused-silica block, and replacing anodic bonding with Cu-Cu thermo-compression to limit the release of oxygen \cite{Karlen2020SealingBonding}.  A recent vapor cell for atomic strontium \cite{Pate2023} reduces the technological gap for warm vapor quantum sensors based on alkaline earth elements (Sec.~\ref{sec:species}).

One of the challenges remains availability and mass fabrication, so an extremely promising technique demonstrates wafer-scale production of identical vapor cells using a commercial anodic bonding machine \cite{Bopp2020}.  These wafer bonding techniques naturally lead to cells with optical access on two opposing faces.  There remain attractive sensor architectures where stemless vapor cells with optical access all faces would be attractive.  Continued developments in miniature vapor cells to meet the wide needs of the various sensors will support their maturation.

\section{Illustrative Sensor Examples}\label{sec:examples}

By a considerable margin,  atomic clocks of various designs dominate practical atomic sensors that have been commercialized. Recent examples include the Chip-Scale Atomic Clock \footnote{{h}ttps://www.microchip.com/en-us/product/csac-sa65} and two-photon Rb optical clocks \footnote{{h}ttps://www.infleqtion.com/tiqker}.  In this section, we describe two very different sensors--a commercial developed optical magnetometer array and a developing compact accelerometer based on atom interferometry--to illustrate some of the points concerning atomic sensor development that we have emphasized above.  In both cases, the technological details have largely been published and have benefited from community-wide advances over many years.

%{\Justin draw inspiration from here, but SHORT} The atom interferometer accelerometer under development by the Sandia National Lab represents an intermediate prototype for an inertial navigation application \cite{Lee2022} using the atom interferometer methods described above.  Innovations in sensor subsystems make substantial progress towards a fully-integrated compact device that also performs high-fidelity measurements.  This has been achieved through multi-faceted innovations, published over a decade.  While this work presently exists on the breadboard, it reflects how the complexity of the cold atom system (Sec.~\ref{sec:platform}) can be managed following the development principles espoused in Sec.~\ref{sec:development}.

%All atom-based electromagnetic sensors represent 

%not transmitters, unlike classical antennas, these receivers can be electrically small as $\lambda\gg r_{Bohr}$ - dipole approximation

%Atom-based electromagnetic receivers represent an attractive approach for biomedical \cite{} and communications \cite{} applications with additioanl interest in magnetic-aided inertial navigation \cite{}.  Known frequency shifts such as $H=\vec{d}\cdot\vec{E}$ or $H=-\vec{\mu}\cdot\vec{B}$ interactions\\
%All receivers, the RF to photon generation is a huge energy cost\\
%unlike classical dipole antennas, atoms can be electrically small as even optical photons $\lambda\gg r_{Bohr}$\\
%vector projections or scalar value.  Ultimately, there are three degrees of freedom in the vector.  The scalar 
%heading error and dead zones

\subsection{Optical Magnetometer Array}\label{sec:magnetometer}

\begin{figure}
    \centering
    \includegraphics[width=0.85\linewidth]{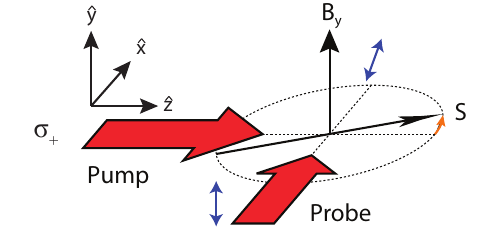}
    \caption{Single component low-field vector magnetometer.  {\Edit Circularly-polarized ``$\sigma_+$" light optically pumps atoms along the z-direction.} A small transverse magnetic field {\Edit $B_y$} rotates the spin {\Edit $\vec{S}$} away from the optical pumping direction, and causes a spin-dependent index of refraction that rotates the plane of polarization of the probe light.}
    \label{fig:mag}
\end{figure}

There are a variety of types of optically pumped magnetometers.  A common version uses modulated light or applied oscillating fields to drive spin precession  at the Larmor frequency $f=|\gamma B|$.  The precession may be simply detected by modulated absorption of the pumping light.  More complex arrangements might instead observe the Faraday rotation of an off-resonant linearly polarized probe laser.  However the detection is done, the frequency can be measured with great precision and the magnetic field magnitude so deduced.  The measurement precision is fundamentally limited by  transverse spin-relaxation, with  spin-exchange collisions dominating when the vapor density exceeds typically {\Edit $10^{11}/$cm$^3$}.  The accuracy of the field measurement is typically limited by non-linear Zeeman and AC-Stark effects.  Such magnetometers have ``dead zones"--orientations of the magnetic field relative to the sensor where low precession amplitudes degrade the sensitivity.  With these caveats in mind, such magnetometers typically have pT scale accuracy and sensitivity, though their QPN limits are much lower than this.

In a spin-exchange-relaxation-free magnetometer, optical pumping along direction $\bf k$ spin-polarizes a hot alkali-metal vapor to a modest $\sim 50\%$ spin-polarization in a small (nT or less) magnetic field, $\expect{\bf S}\cdot{\bf \hat k}\sim 1/4$.  The torque from a small transverse magnetic field causes a tilt of the polarization $\expect{\delta \bf S}=2\mu_B {\bf B}\times \expect{\bf S}/\Gamma_2$ out of the $k-B$ plane (Fig.~\ref{fig:mag}).  Spin-exchange collisions, though they scramble the spin of individual atoms, nevertheless conserve total angular momentum and so do not contribute to $\Gamma_2$ when the magnetic field is near zero, see Sec.~\ref{sec:decoherence}.  The transverse relaxation is then determined by a combination of other collision mechanisms and light scattering.  For greatest sensitivity, the transverse spin-polarization is detected by Faraday rotation of a linearly polarized off-resonant probe, and for optically thick samples this detection is effectively of the QND type.

  The quantum projection noise for QND spin detection is limited at high densities by alkali-alkali spin-axis relaxation {\Edit (rate coefficient $\beta_{SA}$)} \cite{Kadlecek1998}, to be
$\delta B_{SE}=\sqrt{\beta_{SA}/V}/\gamma\sim 10$ ${\rm  aT}/\sqrt{\rm Hz}$
for a 1-\strike{cc}{\Edit cm$^3$} volume $V$ of \strike{$K$} {\Edit potassium}.  Such a low noise floor, orders of magnitude smaller than the background noise in even a high quality magnetic shield, implies that there is a substantial margin available to simplify and miniaturize SERF sensors without degrading performance.

\begin{figure*}[hbt]
    \centering
    \includegraphics[scale=0.25]{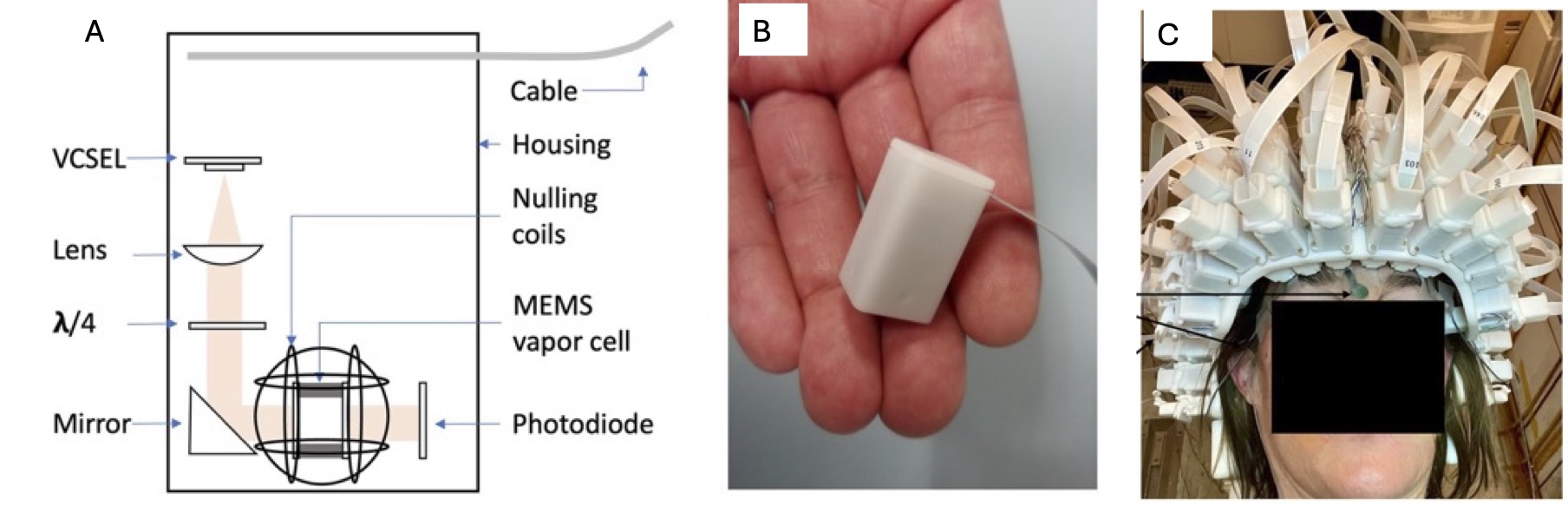}
    \caption{Atomic magnetometer array, adapted from Ref. \cite{Alem2023}. Laser, optics, cell, magnetic field shim coils, and photodiode are integrated inside a 1.5 cc volume (b).  The cable carries electrical signals in/out, along with fiber optic for optical heating of the cell.  128 sensor elements are placed for ultrasensitive magneto-encephalography.}
    \label{fig:Olem2023fig}
\end{figure*}

Alem {\it et al.} \cite{Alem2023} present an instructive recent implementation of an array of 128 high sensitivity optically pumped magnetometers for magnetoencephalography.  The design decisions that are made to implement such a system illustrate many of the principles we have been discussing.

The individual magnetometer elements are microfabricated \cite{Alem2017} glass cells containing Rb vapor in a N$_2$ buffer gas. Each cell is individually heated to about 150 C by an non-resonant laser coupled to the cell through an optical fiber\cite{Alem2017}.  A single VCSEL laser at 795 nm provides the optical pumping, and magnetometer detection is done by measuring the transmission of the VCSEL through the cell. A ``strong" oscillating magnetic field (for a SERF, 150 nT at 1.7 kHz!) along $\hat{x}$, in concert with DC optical pumping along $\hat{z}$ at zero field, produces an oscillating spin component along the $\hat{y}$ direction. Then a small magnetic field in the $\hat{x}$ direction tilts this oscillation into the $\hat{z}$ direction which modulates the transmission of the pumping light.  Laser and field control signals, plus low-noise photodiode readout, are coupled to each element by cabling.  An automated control system optimizes the individual elements, and feeds back to hold the magnetic field zero.  The single channel sensitivity is 20 fT/$\sqrt{\rm Hz}$.

The reader should not be fooled by the apparent simplicity of Fig.~\ref{fig:Olem2023fig}.  Each part of the magnetometer array represents a deep understanding of SERF magnetometry and honing of techniques to simplify it relative to a high performance laboratory apparatus.  Developments of very sophisticated microfabricated cells \cite{Kitching2018} and VCSEL lasers, not to mention heating innovations \cite{Mhaskar2012} and low-noise electronics, are each crucial to integration into a working sensor array.  Sensor generated magnetic fields are managed to limit interference the vapor cell and suppress sensor cross-talk. Even though the sensor does not run at the quantum projection noise limit, the extremely low QPN limit provides a margin that can be exploited for simplicity and scalability that is essential for a practical sensor.  Thus the more sensitive but complex QND Faraday probe detection is purposefully replaced by a less efficient modulated pump transmission method without significantly compromising the necessary performance.  It is also interesting to note that a substantial fraction of the single sensor element volume is occupied by the collimation optics; mode converter technology could potentially allow for the essential physics package (cell plus coils) to occupy a greater fraction of the sensor volume.

\subsection{Cold Atom Accelerometer} 
%Here, $\hbar k_\mathit{eff}$ represents discrete momentum states that can be precisely manipulated through the atomic recoil momentum, though time between pulses $T$ (and transverse velocity $v$ in gyroscopes) must be defined by the instrumentation.  Note, the initial velocity of the atom does not directly contribute to the measurement in a closed atom interferometer \cite{Ben-Aicha2024}.

\begin{figure}
    \centering
    \includegraphics[width=1\linewidth]{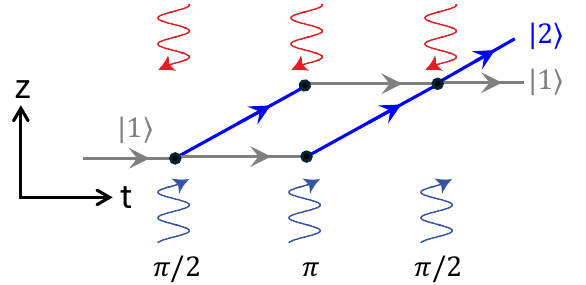}
    \caption{LPAI configured as an accelerometer in a $\pi/2-\pi-\pi/2$ configuration based on Raman transitions that provide electronic state labeling of the output momentum states.}
    \label{fig:ai_figure}
\end{figure}

As the name suggests, the light-pulse atom interferometer uses controlled light pulses to manipulate the atom wavefunction that produce the \strike{the} analogs of the beamsplitters and mirrors in an optical interferometer (Fig.~\ref{fig:ai_figure}).  The pulse timing $T$ and photon momentum $\hbar k_\mathit{eff}$ exchanged with the atom define distinct spacetime trajectories along the interferometer arms.  Transitions driven between momentum states with 50\% probability form a beamsplitter (a $\pi/2$-pulse) to separate/combine atomic trajectories while transitions with a 100\% probability form a mirror (a $\pi$-pulse) to redirect paths.  Raman transitions, Bragg diffraction, and Bloch oscillations offer mechanisms to manipulate momentum states where the Raman transitions offer a convenient electronic state labeling entangled with the output momentum states (Fig.~\ref{fig:ai_figure}).

In the atom interferometer, each atom interferes with itself, encoding the phase in the population of the output momentum states.  As the coherent overlap of the atom wavefunction contributes to the interferometer contrast $C$, laser cooling greatly enhances the atom interferometer performance \footnote{For example, the de Broglie wavelength of room temperature thermal atoms $\lambda_{dB}\sim20$~pm can be extended to $\lambda_{dB}\sim 20$~nm for laser-cooled atoms following the inverse proportionality to the momentum (square root of the kinetic energy) $\lambda_{dB}=h/p=h/\sqrt{2mE}$.}. The measured phase of the atom interferometer can be understood through the formalism of the Feynman Path Integral approach where the phase $\phi$ of the matter wavefunction $\psi$ is given by $|\psi(t)\rangle=e^{-i(S_\Gamma/\hbar)}|\psi(t=0)\rangle$ {\Edit where $S_\Gamma$ is the classical action along path $\Gamma$}.  Changes in energy and interactions along the interferometer arms impact the phase accumulation.  

An insightful interpretation recognizes the standing waves in the optical pulses as ticks on a ruler to provide precise positioning along atomic trajectories \cite{Overstreet2021PhysicallyInterferometry}.  The leading inertial contributions to the phase include both accelerations $\vec{a}$ and rotations $\vec{\Omega}$,
\begin{equation}
\Delta\phi=\vec{k}_\mathit{eff}\cdot[\vec{a}-2(\vec{\Omega}\times\vec{v})]T^2=\vec{k}_\mathit{eff}\cdot\vec{a}T^2+\frac{2m}{\hbar}(\vec{A}\cdot\vec{\Omega})\label{eqn:ai}
\end{equation}
where the $k_\mathit{eff}$ is the effective 2-photon wavevector and $T$ is the time between pulses.
For linear separation of the atoms in a Mach-Zehnder interferometer (Fig.~\ref{fig:ai_figure}), the inertial phase is $\Delta\phi=k_\mathit{eff}{\Edit a}T^2$ where the Coriolis term $-2(\vec{\Omega}\times\vec{v})$ is an important systematic effect \cite{Lan2012InfluenceInterferometry}.  For interferometers that enclose a spatial area $\vec{A}$, both accelerometer and gyroscope terms persist \footnote{Two simultaneous, counter-propagating interferometers with overlapping trajectories can separate the acceleration and rotation by computing $\Delta \phi_1+\Delta\phi_2$ and $\Delta \phi_1-\Delta\phi_2$.}.  The scale factor increases with the enclosed area (spacetime and/or spatial), growing with $T^2$. In all cases, variations in the laser phase stability over $2T$ contributes an important technical contribution to the achieved resolution.

It is important to note that the repeating pattern of fringes provides high-sensitivity to parameters that determine the phase.  As with all interference measurements, sensitivity is best in the vicinity of the fringe crossing and suffers from a limited dynamic range of $\Delta\phi=\pi$, defining $a_\pi$ (or $\Omega_\pi$).  Dynamic range can be extended if fringe disambiguation can be achieved through a simultaneous coarse measurement \cite{Lautier2014HybridizingAccelerometers}. In addition, the scalar phase measurement implies that a single measurement can only produce a projection of the inertial input as indicated by the inner products in Eqn.~\ref{eqn:ai}.

\begin{figure*}[hbt]
    \centering
    \includegraphics[width=1\linewidth]{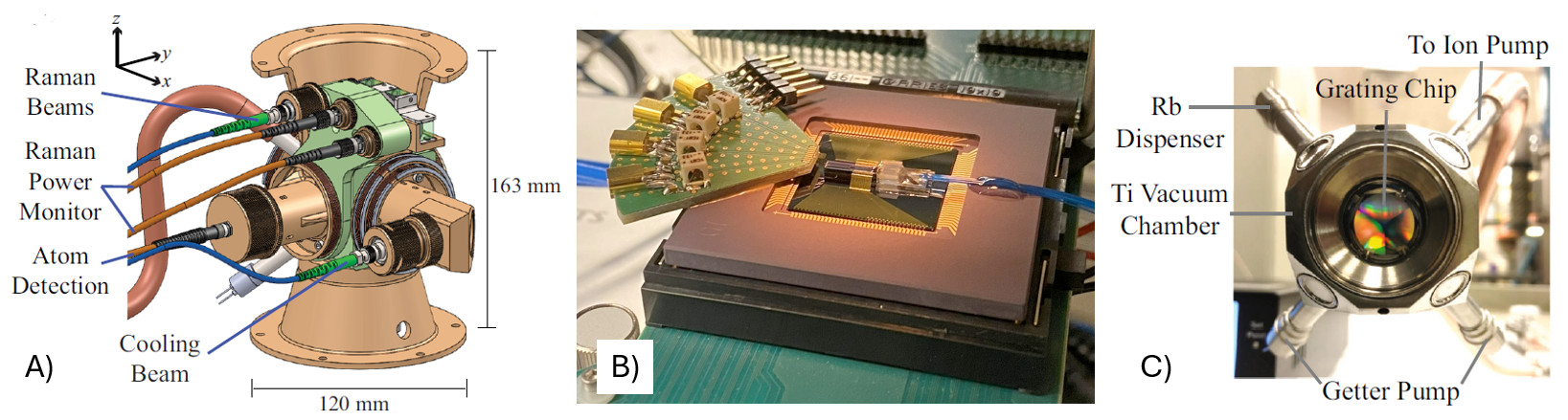}
    \caption{A) Rendering of compact sensor head including optical delivery beams B) Closeup of SSB integrated photonic chip with PCB interconnects that forms the backbone of the optical frequency generation and control system C) closeup of vacuum package highlighting atom source, grating MOT, and spatially separated ion pump through copper umbilical. All images adapted from Ref.~\cite{Lee2022}.}
    \label{fig:sandia}
\end{figure*}

\paragraph{Key Features} Several innovations are highlighted to achieve a compact \strike{baseball}{\Edit softball}-scale sensor head with simplified optical delivery (Fig.~\ref{fig:sandia}a).  Two single-mode fibers deliver light to the tethered sensor head where a single laser beam supports atom trapping through a grating magneto-optical trap (GMOT) while counter-propagating laser beams that drive two-photon Raman transitions are delivered by a separate fiber.  Simplified, high-rate atom loading occurs through an atom recapture method compatible with atom interferometry \cite{McGuinness2012}.  Additional fibers collect light for state detection and diagnostic monitoring through multi-mode optical fibers to limit interference from electrical signals on the atoms.  In the same vein, rigid, non-conducting materials support the structure for robustness and limit eddy currents under magnetic field switching at $\sim100$~Hz.  Fixed optics (beamsplitters, polarizers, mirrors, waveplates) used throughout the sensor head remove mechanical degrees of freedom for long-term stability where measurements are enabled though time-multiplexed frequency shifting from a single seed laser.

\paragraph{Optical Frequency Generation} To address the challenge of bulky atom interferometer laser systems, \strike{the authors establish }a technical development path towards a PIC based on existing silicon photonics at 1560~nm {\Edit is established}, taking advantage of maturity of the telecom-supported photonic infrastructure and frequency-doubling to the $^{87}$Rb D2 line at 780~nm.  As a PIC offers minimal adjustment post fabrication, a rack-mounted version {\Edit of} the laser system with fiber-connected {\Edit commercial-off-the-shelf}\strike{COTS} components {\Edit (akin to Sec.~\ref{sec:breadboarding})} supports the measurements and validates the optical configuration {\Edit integrated into the PIC} \cite{Kodigala2024High-performanceInterferometry}. Figure~\ref{fig:sandia}b showcases the workhorse PIC component developed internally, a single-sideband modulator (SSB), enabling rapid control of a second co-propagating frequency by adjusting the RF input.   Common phase and polarization of the carrier and sideband builds on previous demonstrations with discrete SSB modulators \cite{Zhu2018SSB,Rammeloo2020}. The PIC achieves close to single frequency generation (30 dB carrier suppression and 47.8 dB sideband suppression)  to reduce systematic effects from lightshifts \cite{Carraz2012PhaseModulation}.  A similar system has been developed in Ref.~\cite{Templier2021}.

\paragraph{Atomic Vapor Isolation} A low out-gassing, helium impermeable vacuum chamber made from titanium with five anti-reflection coated fused silica windows \strike{encloses} {\Edit encapsulates} the {\Edit enclosed} measurement volume (Fig.~\ref{fig:sandia}c).  Compared to traditional steel chambers, the titanium offers lower hydrogen offgassing and lower magnetization.  Here, an ion pump connected via copper umbilical reduces the impact of magnetic field gradients on the measurement while maintaining UHV.  A parallel development using a similar titanium chamber with helium-impermeable sapphire windows eliminates the pump entirely  \cite{Little2021}.  Atom loading is accomplished through rapid recapture {\Edit of} the MOT{\Edit .}\strike{,} {\Edit This technique} demonstrat{\Edit ed}\strike{ing} $>80\%$ recapture efficiency \strike{first demonstrated with rapid atom interferometry} in Ref.~\cite{McGuinness2012}{\Edit , but suffers from reduced efficiency in the GMOT due to a lower capture volume}.

\paragraph{Measurement Considerations} The prototype takes an intermediate advancement on delivery of light using a racetrack configuration.  As discussed previously, the laser phase between the two Raman frequencies represents a significant noise contribution, so common optical path with retroreflection mirror would enable the Doppler-sensitive Raman transitions with counter-propagating photons.  However, this configuration leads to degeneracy at zero center of mass velocity of the cloud between the forward-propagating and reflected optical frequencies \cite{Tino2020}, limiting operation when no acceleration is present, a poor match for an inertial navigation application.  The racetrack splits orthogonal linear polarizations delivered through a common polarization maintaining fiber to enter from opposite sides of the chamber.  Recent publications have introduced zero-velocity compatible methods with fixed optics such as a frequency chirp \cite{Perrin2019} or an symmetry-breaking pair of magnetic insensitive transitions from the clock state \cite{Bernard2022}.  Other techniques such as adiabatic rapid passage \cite{Kotru2014} and top hat beams \cite{Mielec2018} address cross-axis accelerations. Single-axis measurements in this system can be extended to 3-axis measurements as has been demonstrated in Ref.~\cite{Templier2022TrackingTriad}.

\paragraph{Precision \& Accuracy}  The rapidly repeating measurements provide a statistical sensitivity in the measured acceleration of 
\begin{equation}
\sigma_g=\frac{\delta \phi}{C k_\mathit{eff}T^2\sqrt{R}}
\end{equation}
where $\delta\phi$ is the single shot uncertainty in $\Delta \phi$ and $R$ is the cycle rate.  For duty cycles dominated by measurement time where $1/R\sim T$, this leads to a $1/T^{3/2}$ scaling in the measurement precision.  Early breadboard measurements demonstrated  $R=50-330$~Hz achieving $0.57~\mu g/\sqrt{{\rm Hz}}$ and $36.7~\mu g/\sqrt{{\rm Hz}}$ respectively \cite{McGuinness2012}.  These measurements were limited by technical noise, laser phase noise of 21~mrad/shot attributed to optical component instability, and magnetic field noise of 15~mrad/shot, substantially larger than the QPN phase $\delta\phi_{QPN}=1/\sqrt{N}<1$~mrad/shot for $10^6-10^7$ atoms.  {\Edit When vertically-oriented,} \strike{M}{\Edit m}easurements in the compact sensor head exhibit a slowly varying offset which can be eliminated by fitting to a chirped sinusoid to measure local gravity to a statistical precision of $\Delta g/g=2.0\times10^{-6}$.  An offset of $1.6$~mm/s$^2$ from the nearest surveyed gravity value (adjusted for elevation) indicates introduction of a systematic effect not present in the initial demonstration, highlighting the continuing value of the breadboard experiments (Sec.~\ref{sec:breadboarding}).

\section{Advanced Quantum Methods}\label{sec:QE}

Advanced quantum methods, such as squeezing (light and/or spins) and entanglement, allow for measurements with statistical precision below the standard quantum limit. Squeezing reduces quantum noise in one observable by increasing it in another conjugate observable.  Entanglement exploits quantum correlations for increased sensitivity, and for certain types of non-local states allows the quantum noise to scale as $1/N$ as opposed to $1/\sqrt{N}$ for the standard quantum limit \cite{Holland1993InterferometricLimit, Giovannetti2006QuantumMetrology}.  Generally, such advanced quantum methods allow improvements in statistical uncertainties and increase bandwidth.  
They are therefore of fundamental interest to quantum sensors of all types.

The impact of squeezing methods can generally be quantified by the Wineland parameter $\xi$ as a prefactor in Eqn.~\ref{eqn:one} where $\xi \le 1$ \cite{Wineland1994,Szigeti2021}.  For a fixed number of atoms, a quantum enhanced measurement offers increased statistical resolution per measurement by $\xi$, assuming of course that non-quantum noise sources have been eliminated.
With this assumption, $\xi < 1$   decreases the short-term noise in the Allan deviation plot (Fig.~\ref{fig:allan_deviation}), with little impact on the bias stability \cite{Schulte2020}.   Some types of entangled states  exhibit Heisenberg scaling Ref.~\cite{Huang2024}
 (${1/\sqrt{N}}\rightarrow  \sqrt{N_0}/N$) which  for $N>N_0$ would lead to even faster averaging to the bias stability limit.

The most dramatic and impactful use of advanced quantum methods for metrological use is undoubtedly the implementation of frequency-dependent squeezing in the LIGO gravitational wave detector \cite{Ganapathy2023}.   Another beautiful exploitation of advanced quantum methods was the recent dark matter search using squeezed microwaves \cite{Backes2021}.  While neither of these are atomic sensors, they illustrate the potential advanced quantum techniques have for metrological gain.  In both of these cases, the experimental sensitivity due to classical noise sources had been pushed below the SQL, 
making quantum metrology a necessity to improve the sensing performance.

For atomic sensing, proof-of-principle experiments have shown impressive improvements of measurement phase sensitivity below the SQL of $10$~dB in Ref.~\cite{Bohnet2014ReducedLimit} and $20$~dB in Ref.~\cite{Hosten2016MeasurementAtoms}.  Recent demonstrations of metrologically useful gains include clocks \cite{Pedrozo-Penafiel2020EntanglementTransition,
%Bornet2023, - not sure how I got here
Eckner2023RealizingClock},  magnetometers %\cite{Baranga2024} Baranga2024 is a less clear correlated atom result
\cite{Troullinou2023Quantum-EnhancedDensity,Zheng2023Entanglement-EnhancedTomography}, and atom interferometers \cite{Anders2021}. These experiments used spin-squeezing or squeezed light to attain 2-5 dB performance beyond the SQL.  They also correspond to sensor configurations where atom number cannot be readily increased without increasing collisional shifts (clocks) or shortening decoherence times (rf magnetometers).  As such, these experiments represent important milestones toward implementing advanced quantum methods in practical quantum sensors.

\begin{figure}
    \centering
    \includegraphics[width=1\linewidth]{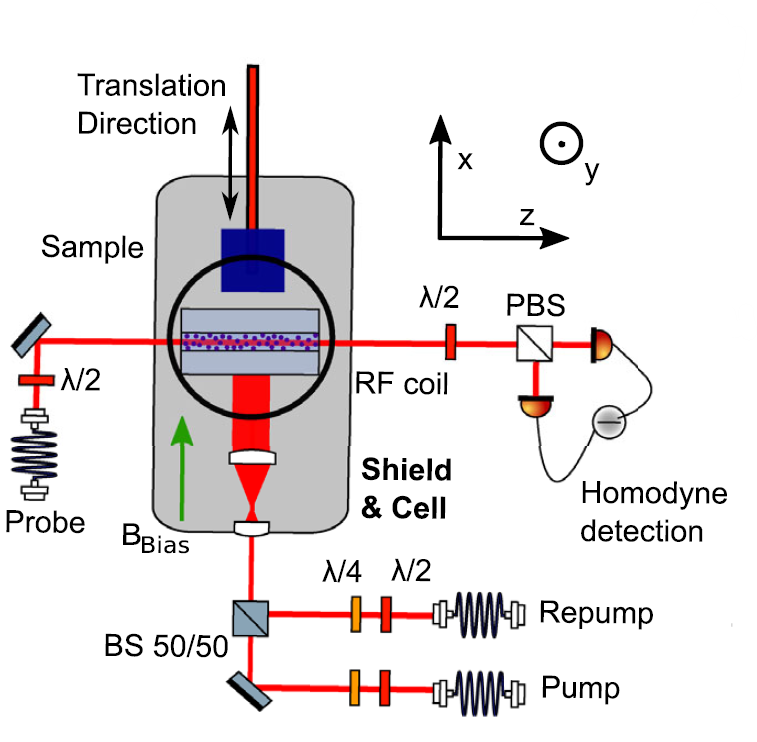}
    \caption{Apparatus for enhanced RF magnetometry using spin-squeezing {\Edit not showing AOMs on pump and probe}. Adapted from Ref.~\cite{Zheng2023Entanglement-EnhancedTomography}. }
    \label{fig:Polzik}
\end{figure}

An example of a spin-squeezed rf magnetometer for sub-SQL detection of magnetic induction tomography \cite{Zheng2023Entanglement-EnhancedTomography} is shown in  Fig.~\ref{fig:Polzik}. The 4.6 dB of squeezing is acheived using a stroboscopic probing scheme with a pair of QND spin-detections.  The complexity cost of this implementation of spin squeezing is notably small and with PIC technology could be readily incorporated into arrays.  This illustrates that electromagnetic sensing, which cares primarily about short term sensitivity, is a promising avenue for implementation of advanced quantum methods.  

A high complexity-cost example of high performance quantum sensing is a spin-squeezed optical clock using Rydberg interactions in a Sr atom array \cite{Eckner2023RealizingClock}.  As always appropriate for clocks, the authors stress
the importance of assessing the impact of advanced measurement methods on systematic errors \cite{Schulte2020}; this issue is not always addressed explicitly.

With some exceptions, advanced quantum methods to date have been laboratory implementations where the additional complexity of these methods can be managed by experts.  For compact, ``practical" atomic sensors,   the increased infrastructure and hence complexity cost of most metrologically useful advanced quantum demonstrations to date is too great to be currently feasible \cite{Szigeti2021}. Research leading to simplification of advanced quantum methods is key to their implementation in {\Edit practical} atomic sensors.

\section{\label{sec:open}Open Challenges and Future Developments}

The overall challenge facing the development of atom-based quantum sensors is the process of transitioning the four ``I"s of Sec.~\ref{sec:atom_adv} (\strike{Indistinguishable} {\Edit Identical}, Isolate-able, Interface-able, \& Intelligible) from the laboratory to the field.  A balance must be struck to maintain as much as possible of the high-value sensitivity and stability of laboratory experiments, but with proper regard to practical issues (SWaP, ruggedization, manufacturability, scalability etc.).  Continued innovations in sensor designs and supporting technologies as well as advances in fundamental understanding will allow modern atomic quantum sensors to manage both complexity and performance which can enable widespread implementation.

We conclude by offering some observations about additional developments and particular topics for further emphasis in atomic sensor research.

\subsection{Emergence from Fundamental Research}

To our knowledge, all current atomic sensor technologies are outgrowths of fundamental, curiosity-driven research.  Complex quantum sensing apparatus, developed for expert use to address scientific questions, was recognized to have potential to be adapted for practical quantum sensors.  New techniques in manipulation and measurements on atoms, coupled with technological advances from other fields, are being translated into compact, practical quantum sensors.  The contributors to Physical Review can assert, without irony, that their fundamental research--however abstruse--has long-term broader societal impacts.

We have argued that the process of translation from sensing to sensor is best accomplished in stages.  As prototypes are built and tested, it is critical to ``bread-board" for the purposes of taking advantage of fully instrumented diagnostics to make key studies of the fundamentals of the sensor. To proceed without this step overlooks valuable opportunities for innovation.  Understanding the nature of systematic errors and noise sources is critical to intelligent implementation of any technology in the field.  Likewise, understanding fundamentals of decoherence mechanisms is essential to sensor design and implementation, but often difficult to ascertain and understand with limited diagnostics of matured sensor packages.  It is here that the breadboard maintains its value throughout the sensor development process.

Although all types of AMO research can and do contribute to the development of atomic sensors, precision measurement studies are particularly well-suited to sensor spin-offs. One can view a practical atomic sensor as a portable precision measurement device,  designed to allow an unskilled operator to produce an accurate measurement of the desired quantity under a variety of environmental conditions. For both the sensor and the lab experiment, a device is designed to be maximally sensitive to the quantity of interest and reject other confounding variables.  Due to the extreme sensitivity sought, measurement simplicity is valued and preserved for purposes of intelligibility and complexity management.  An {\Edit excellent} \strike{interesting} example \strike{are} {\Edit is} the continuing search\strike{es} for {\Edit a} permanent electric dipole moment\strike{s} of elementary particles.  Competitive limits on  permanent electric dipole moments continue to be set with vapor cells \cite{Griffith2009,Graner2016} and beam systems \cite{Hudson2011,Andreev2018}, as well as ion traps \cite{Cairncross2017}.  As these techniques are stretched to their limits,  new efforts \footnote{ACMEIII Collaboration: https://cfp.physics.northwestern.edu/gabrielse-group/acme-electron-edm.html} are incorporating laser cooling and trapping techniques.  Lessons learned in management of complexity cost in state-of-the-art experiments like these will doubtless translate into new insights in practical quantum sensors.

Thus, the notion that the transition from sensing to sensor is ``just an engineering problem" is naive and will likely lead to underperformance; new issues arise in sensors that were not addressed in their laboratory forebears.  In addition, the progenitor experiments were optimized for different environments or purposes, requiring new variations on the established theme to be developed.  Instead, atomic sensor development should be recognized and appreciated as an independent intellectual endeavor. 

\subsection{Publication of Atomic Sensor R\&D\label{sec:publication}}

The common attitude that quantum sensor development is primarily an exercise in engineering has other unfortunate side effects.  In our experience, this view has led both sensor developers and funding sources to devalue disseminating information about the sensors.  Of course, there is an understandable reluctance on the part of industrial firms to share their hard-earned ``secrets" that were possibly developed at high internal cost;  however, in the modern era, many if not most atomic sensor development projects are actually funded by government sources, a cost largely borne by the public.  The same pressures that are driving open source publication of publicly funded fundamental research should apply to public funding of R\&D as well. Suppressing the dissemination of details of quantum sensor achievements slows overall development, also not in the public interest.

%Suppressing the dissemination of details of quantum sensor developments unfortunately slows development. It also  withholds proper recognition of the creativity and accomplishments of the organizations and dedicated scientists who advance the field.  These advancements should be submitted to leading journals for dissemination, rather than limited to conference proceedings. {\Justin Help me restate} If it isn't communicated, it is equivalent to not having been achieved.

%By its very nature, fundamental research at the forefront of knowledge forces researchers to develop new techniques and manage complexity.  All the examples of sensors we have encountered grew out of complex laboratory apparatus developed to address scientific questions, not advance measurement capabilities per se.  Even for research directly interested in sensor development, we have argued that is it is critical to ``bread-board" more complex sensor versions for the purposes of taking advantage of fully instrumented diagnostics to make key studies of the fundamentals of the sensor.  

\subsection{Physics in Challenging Environments}
Research done outside the relatively pristine, well-regulated environments of modern  laboratories at universities and government or industrial research institutes naturally shares the goals of practical quantum atomic sensors.  Space and budget constraints, plus the need for the experiments to often operate remotely or survive large temperature variations, platform dynamics, or operation by non-specialists, drive the development of simpler, more robust techniques that are synergistic with improved quantum sensors.  ``Crazy" ideas, pursued to satisfy curiosity, inspire advances in methods, capabilities, and understanding that will eventually be incorporated in quantum sensors.

\subsection{Hot Complements Cold}

We have emphasized in this article the complementary nature of cold and hot atom technologies.  When possible, the simplicity and compactness of sealed vapor cells make hot atoms a clear choice.  For magnetometry, the vastly greater number of atoms with comparable decoherence rates make hot atoms vastly superior.  For compact clocks of not too high performance, one is often willing to live with greater decoherence with hot atoms in order to gain in simplicity, as in the CSAC or 2-photon clock.  For high performance clocks, the cost of greater SWAP and complexity with cold atoms is more than compensated by superior sensitivity for the most demanding applications.   For accelerometers, cold atoms are essential, as the momentum state preparation by laser cooling is the only viable approach. Atomic rotation sensing can be done with either hot atoms (NMR gyroscopes \cite{Wei2023Comagnetometer} ) or with interferometry\cite{Narducci2022}, each with their advantages and disadvantages, meriting research focus.

\subsection{Supporting Technology Infrastructure}

%We cannot overemphasize the importance of  s
Specialized techniques and tools are essential for advancing atomic quantum sensors.  Leading the way here  are improvements in laser technologies, addressing performance, cost, and simplicity.  Improved laser sources and techniques to stabilize them will almost always find their way into atomic sensors.  We are particularly excited about new possibilities being enabled by rapidly advancing PIC technology.  Perhaps less glamorous but no less important are advances in electronics, detectors, optical components, and vacuum systems that are present in every atomic quantum sensor.  Indeed, an incredibly important issue facing wide implementation of cold atom technology is the relatively bulky and intrusive nature of vacuum pumps.  Development of compact, field-free, long-lived vacuum systems compatible with minimizing decoherence in cold atom systems would greatly enhance the ways that cold atoms could be implemented as sensors.  Similarly, developments of more consistent, mass-producible methods of custom vapor cell manufacturing are key to hot atom sensor arrays.

\subsection{Quantum Measurement}

Advances in quantum measurement techniques have the potential to improve the performance of atomic sensors, particularly in reducing the time required to reach bias stability limits.    With few exceptions, implementation of advanced quantum measurement methods has generally to date been accompanied by a substantial increase in complexity cost that, in the short term, inhibits their use in practical sensors.  Research and development that emphasizes both enhanced performance and simplification of advanced quantum measurement techniques is essential to advance state-of-the-art atomic sensors.

%Any quantum sensor, and indeed any fundamental science experiment, must focus on both statistical and systematic errors.  The best sensors strike a balance between both.  In practice, fluctuating systematics often form a noise floor that keep sensors from reaching the SQL, a necessary requirement before advanced quantum measurement methods even become an option.

%As exciting as quantum measurement (entanglement, squeezing) are from a fundamental science perspective, the fragility of the techniques remain difficult to implement in harsh environments or  geometries of interest for compact atomic sensors. Development of robust methods will be critical.

\subsection{Quantum Information Scale-Up}
%Scaling up of digital quantum information processing technology}

Digital quantum information processors, which for atoms take the form of tweezer arrays or chains of ions in traps, are in one respect assemblies of single-atom quantum sensors.  They have been elegantly designed to avoid decoherence as much as possible, and are ideally suited for implementing advanced quantum measurement techniques that exploit entanglement.  With the exception of ion clocks, the limited numbers of qubits to date have generally made sensors based on qubit technologies uncompetitive with other approaches.  As we look to the future, the number of qubits that can be usefully manipulated is rapidly increasing with time, and at some point the implementation of Quantum Information Processing (QIP) systems may become practical.  {\Edit Quantum computers are already being proposed as methods to improve the sensitivity of arrays of quantum sensors \cite{Allen2025arXiv}.}

An intriguing question involves possible generalization of error correction techniques to quantum sensors.  We have pointed out that quantum measurement methods such as squeezing usually help with statistical errors but are still subject to decoherence issues.  Error correction is essentially the preservation of coherence/entanglement in the presence of decoherence.  Can the concept of the logical qubit be extended to metrologically interesting and important sensors?  Can error correction techniques reduce the sensitivity of sensors to systematic errors?  Likely these exciting prospects will first occur in the context of clocks, but could be important for other types of sensors as well.

\subsection{Complex Atoms and Molecules}

Many of the exciting advances in AMO physics over the past decade have come in the rise to prominence of alkaline-earth-like atoms and molecules.  The greater complexity of these species in the past made them impractical, but, building on a recurring theme of this article, laser advances coupled with improvements in fundamental understanding of these systems have led to remarkable achievements in laser cooling, coherent manipulation, and fundamental studies of their properties.  As the infrastructure involved in such experiments steadily simplifies, capabilities not possible with traditional alkali-metal atoms may be accessible for sensors. {\Edit For example, continuous BEC loading has been demonstrated in Sr \cite{Chen2022ContinuousCondensation}.} 

Laser-cooled molecules represent an emerging platform for quantum simulation, quantum information processing, and tests of fundamental physics that could evolve into specialized sensors.  The additional rotational and vibrational states of even a diatomic molecule add substantial complexity relative to most atoms, especially in the number of required laser sources, but enable unique measurement modalities.  Do the benefits of molecules justify the complexity cost?  Can the techniques be simplified \cite{Li2024}?

\begin{acknowledgments}

T.W. acknowledges support from the National Science Foundation, grant \# PHY-1912543.

\end{acknowledgments}

\appendix*
\section{Additional Sensor Resources\label{app:resources}}
 Some helpful, recently-published resources in support of atomic sensors, organized by topic, include:

\paragraph{Atomic Sensors} A few resources exist to broadly address atom{\Edit -}based sensors including a review of atomic sensors \cite{Kitching2011} and chip-scale atomic sensors \cite{Kitching2018,Yu2024KeyMetrology}.  Several perspectives are available on quantum sensors \strike{that includes} that emphasize atom-based sensors for quantum information science \cite{Metcalfe2023}, basic research to commercial applications \cite{Oh2024PerspectiveApplications}, and deployment of quantum sensors \cite{Bongs2023}.  {\Edit Also relevant is a 2019} \strike{The} Defense Science Board {\Edit report} \strike{advocates for continued development of quantum clocks and accelerometers for defense applications and is less keen on atom interferometer gyroscopes} \cite{DSB2019}.

\paragraph{Clocks} Recent reviews focus on optical clocks \cite{Ludlow2015}, vapor cell frequency standards \cite{Godone2015High-performingStandards}, optical frequency combs \cite{Diddams2020}, commercial frequency standards \cite{Marlow2021AStandards}, and more broadly atomic frequency standards \cite{Vanier2016TheStandards}.  Nuanced metrics for practical operation outside the laboratory are detailed in the {\it IEEE Guide for Measurement of Environmental Sensitivities of Standard Frequency Generators} \cite{20231193-2022Generators}.  Reviews on transportable optical clocks \cite{Gellesch2020} and geodesy applications \cite{Mehlstaubler2018} are also available.

\paragraph{Magnetometers} Introductions to atomic magnetometers range from easily digestible \cite{Budker2007OpticalMagnetometry} to comprehensive \cite{Budker2013OpticalMagnetometry}.  A good tutorial on how to build an atomic magnetometer is available for those interested in laboratory research \cite{Fabricant2023}.  Discussions of magnetometry application in industry  \cite{Bai2023AtomicIndustry} as well as magnetometry for challenging environments \cite{Fu2020} provide helpful context for sensor development.  It may be possible to explore the  quantum limits energy resolution of magnetic field sensors as described in Ref.~\cite{Mitchell2020Colloquium:Sensors}.

\paragraph{Rydberg Sensors} Several reviews cover Rydberg atoms as quantum technology building blocks \cite{Adams2020}, measuring electric fields \cite{Liu2023ElectricAtoms}, and in the context of RF sensing \cite{Holloway2022}.  An overview of microwave sensing \cite{Artusio-Glimpse2022Mag} as well as a comparison to a classical antenna \cite{Botello2022Noise,Weichman2024} provide helpful context for Rydberg sensor development.  Most sensors favor alkali atoms, though opportunities for alkaline-earth atoms are being explored \cite{Dunning2016RecentAtoms}. Rydberg sensing is a rapidly evolving area leading to recent innovations in frequency comb readout and preparation \cite{Dixon2023} as well as sequential application of laser and microwave fields \cite{Romalis2024VaporFields}. % {\Justin Update References \cite{Fancher2021Review, Schlossberger2024Review,Tu2024}}

\paragraph{Atom Interferometers} An accounting of the unexpected inertial sensitivity to the atom interferometer phase emerging from fundamental research serves as a reminder of the value of curiosity-driven research \cite{Chu1998TheParticles}.  A LPAI review \cite{DeAngelis2009PrecisionSensors} as well more comprehensive review including historical development and types of atom interferometers \cite{Cronin2009} provide a helpful introduction to the field.  A detailed description of phase calculations based on atom interferometer geometry remains relevant \cite{Storey1994}.  Several recent reviews highlight specific topics such as gravimeters \cite{Geiger2020High-accuracySensors}, gyroscopes \cite{Alzar2019GyroReview}, and atom interferometers for use outside the laboratory \cite{Narducci2022,Bongs2019}.

\paragraph{Nuclear Spin Gyroscope}\label{app:NMR} Vapor cell gyroscopes can be formed from nuclear spins as either {\Edit a }dual-species xenon isotope driven system \cite{Walker:2016fk} or an alkali-metal noble-gas comagnetometer \cite{Kornack2005NuclearComagnetometer}.  Unlike the atom interferometer, the overlapping spin ensembles have guaranteed overlap\strike{,}{\Edit ;} however, collisions introduce {\Edit difficult to diagnose} long-term stability challenges.  A recent implementation using K-Rb-$^{21}$Ne (developed for fundamental science) demonstrates remarkable advancement when interpreted as a gyroscope{\Edit ,} providing strategic grade sensitivity {\Edit of} $0.0001^\circ/\sqrt{{\rm hr}}$ with a bias stability of 0.004$^\circ/$hr \cite{Wei2023Comagnetometer}.

\paragraph{Supporting Technology}  Some recent reviews \cite{Bogdanov2017MaterialPhotonics,Moody20222022Photonics,Giordani2023IntegratedTechnologies} broadly cover integrated photonics for quantum applications and a more specific review details PIC technologies to create MOTs \cite{Blumenthal2024EnablingApplications}.  A report from the Quantum Economic Development Consortium on Photonic Integrated Circuits for Quantum Applications is available to member organizations here \cite{QEDC-PICS23}.  Comprehensive surveys of miniature cold atom technology \cite{Rushton2014, McGilligan2022} provide useful insight on topics ranging from sourcing atoms, leak mitigation (UHV bonding and coefficient thermal expansion matching), and outgassing (H$_2$, CO, noble gasses etc.) as well as laser frequency generation.  A resource on laser cooling at the  chip-scale is also available \cite{McGilligan2020LaserPlatform}.  Miniature vapor cell construction techniques are thoroughly reviewed by Kitching \cite{Kitching2018}, Knapkewicz \cite{Knapkiewicz2019} and Gorecki {\it et al.} \cite{Gorecki2019}.  A recent review of VCSELs \cite{Pan2024} as well as a list of commercially available VCSELs at relevant atomic wavelengths that specifies vendors \cite{Zhou2022VCSEL} are available.  

\paragraph{Advanced Quantum Measurement}
Ref.~\cite{Schulte2020} provides a detailed analysis of challenges of implementing advanced quantum methods associated with metrological improvements including systematic effects and the Dick Effect for optical clocks.  Deutsch provides an additional perspective on harnessing the Quantum Revolution \cite{Deutsch2020HarnessingRevolution}.  Ye and Zoller provide a relevant essay on Quantum Sensing \cite{Ye2024Essay:Physics}.  {\Edit Several other relevant references include: }\strike{A}{\Edit a} theory paper on number unconstrained quantum sensing \cite{Mitchell2017Number-unconstrainedSensing}\strike{.  Prospects}{\Edit , prospects} for using quantum error correction for improved sensing \cite{Zhou2018AchievingCorrection}\strike{.  Advanced} {\Edit , and advanced }quantum methods without entanglement \cite{Braun2018Quantum-enhancedEntanglement}.

%{\Lingering Quantum Projection Noise in an atomic fountain, PRL \cite{Santarelli1999}}\\

%\paragraph{Sensor Arrays} \cite{Sutter2020, Limes2020} laser noise in clocks \cite{Zheng2022}

\newpage

\bibliography{AtomSens4} % Produces the bibliography via BibTeX.

\end{document}